\PassOptionsToPackage{pdfborder={0 0 .3}}{hyperref}
\documentclass[gmd, hvmath]{copernicus}

\usepackage{booktabs}
\usepackage[version=4]{mhchem}
\usepackage{siunitx}
\DeclareSIUnit\year{yr}
\DeclareSIUnit\month{mo}
\DeclareSIUnit\molar{\mole\per\cubic\deci\metre}
\DeclareSIUnit\equiv{equiv}
\DeclareSIQualifier\carbon{\ce{C}}
\DeclareSIQualifier\silicon{\ce{Si}}
\DeclareSIQualifier\phosphorus{\ce{P}}
\DeclareSIQualifier\nitrogen{\ce{N}}
\DeclareSIQualifier\oxygen{\ce{O_2}}

\DeclareRobustCommand{\DutchName}[4]{#2~#1}

\begin{document}\hack{\sloppy}
\def\introductionexists{true}
\def\conclusionsexists{true}

\nolinenumbers              

\title{Long-term climate simulation in NorESM:\\
       burst-coupling the sediment in the BLOM/iHAMOCC ocean module}
\Author[1,3]{Marco}{van~Hulten}
\Author[1,2,3]{Christoph}{Heinze}
\Author[2,3]{J\"org}{Schwinger}
\Author[2,3]{Jerry}{Tjiputra}
\affil[1]{Geophysical Institute, University of Bergen, All\'egaten~70, 5007 Bergen, Norway}
\affil[2]{NORCE Norwegian Research Centre~AS, 5838 Bergen, Norway}
\affil[3]{Bjerknes Centre for Climate Research, Norway}

\runningtitle{Burst-coupling sediment model in BLOM/iHAMOCC}
\runningauthor{M.M.P.~van~Hulten et~al.}
\correspondence{M.~M.~P.~van~Hulten <marco@hulten.org>}


\firstpage{1}
\maketitle


\hrulefill

\begin{abstract} 
In this report we set forth a simulation method for long-term simulations of
NorESM, the Norwegian Earth System Model.
In this the sediment is repeatedly decoupled and coupled to the ocean model
(BLOM/iHAMOCC), a process called \emph{burst coupling}.
Through this, the ocean (seawater and sediment) is brought into an approximate
steady state.
We show that just the model has to run at least 50\,000\;yr to
get in an approximate steady state.
With burst coupling this can be done in a computationally reasonable time (wall
time in the order of one week).
The method can be used to generate the sediment over hundreds of thousands of
years, so it is useful not only for present-day simulations but also for
paleo-climatological studies.
\end{abstract}

\introduction                           \label{sec:introduction}

The ocean plays a major role in the regulation of the past and present climate.
It has the capacity to absorb huge amounts of heat, as illustrated by the fact
that the ocean was the dominating net heat sink in the Earth system during
1971--2018 \citep{ar6wg1}.
The ocean is also a big net carbon sink \citep{sabine2004,steinfeldt2009}.
The preservation of carbon, both in its organic and inorganic form, affects the
long-term climate \citep[for calcite]{archer1994}.
These effects are long-term, but we need to consider them even for the present
climate.
When we model the present climate, the full physical and biogeochemical
system from upper ocean (fast) down into the sediment (slow) needs to
approximate a steady state.
This can be achieved through model simulation of the ocean (either coupled to
the atmosphere or stand-alone).
The ocean circulation is operating at a time scale of hundreds to thousands of
years.
Reaching an approximate steady state of the physical and biogeochemical system
(complete mixing of the ocean) takes more time.
The sediment is even slower and takes at least tens of thousands of years to
converge to a steady state.
However, using direct model integration, the simulation of the water column is
computationally slower and much more expensive than that of the sediment.
In any case, it is not feasible to do a coupled integration of an ocean model
for tens of thousands of years.

There are several reasons why we are studying the sediment in the
light of understanding and prognosis of the climate.
Firstly, sediment traces climate variations in the past.
Secondly, the sediment is a constraint on the particle fluxes simulated (the
sediment is the best sediment trap).
Thirdly, fluxes between the sediment and the water column are important (oxygen,
alkalinity, carbon, silicic acid and other nutrients).
The \chem{CaCO_3} dissolution feedback is important for fosil fuel \chem{CO_2}
neutralisation. 

A pilot study on the use of sediment observations combined with a modelling
approach (forward and inverse) was published by \citet{heinze2016} using the
simpler, annually averaged HAMOCC2 model.
The model allows for artificial sediment core development in order to 
directly compare model data with sediment core measurements.
In the pilot study, the modelling approach was successful in attributing 
a series of processes as likely candidates for causing the
80--100\;ppm reduction of atmospheric \chem{CO_2} at the last ice~age.

Whereas those simulations were very useful to get insight in the first-order
effects of the processes related to climatic changes at the end of the last ice
age, a state-of-the-art biogeochemical model simulation is useful for a more
realistic analysis.
In this report we set forth a simulation method for long-term simulations of
NorESM, the Norwegian Earth System Model.
In this the sediment is repeatedly decoupled from and coupled to the ocean model
(BLOM/iHAMOCC), a process hereafter called \emph{burst coupling}
\citep{hasselmann1991}.
Through this, the ocean (seawater and sediment) is brought into an approximate
steady state.
We will show how long the system has to run to get in a steady state.
We will show how fast the burst coupling method is; if a long simulation can be
done in a computationally reasonable time.

\section{Methods}             \label{sec:methods}

\subsection{Model description}


We use the isopycnic general circulation model BLOM (Bergen Layered Ocean
Model), which was originally developed from the Miami Isopycnic Coordinate
Ocean Model (MICOM\@) \citep{bentsen2013}.
We use the model within the framework of the Norwegian Earth System Model
(NorESM) \citep{kirkevaag2013}, but in this study the ocean is the only active
(online) component.
Hence this section describes only the ocean model; the other parts of the Earth
system (atmosphere, land and sea ice) do not run on-line in this study but are
data components that force the ocean.

\subsubsection{Seawater processes}
The biogeochemistry is part of BLOM and is referred to as the isopycnic HAMburg
Ocean Carbon Cycle (iHAMOCC\@).
It is a NPZD-type (nutrients--phytoplankton--zooplankton--detritus) model
extended to include Dissolved Organic Carbon (DOC\@).
The model iHAMOCC simulates the cycles of carbon, three major nutrients
(phosphate, nitrate and silica) and the trace nutrient iron, along with
phytoplankton, a zooplankton grazer, detritus or henceforth Particulate Organic
Carbon (\chem{POC}), as well as calcium carbonate (\chem{CaCO_3}) and biogenic silica
(\chem{bSiO_2}).
A constant Redfield ratio of \chem{P:C:N:O_2}~$=$~\(1:122:16:-172\) is used
\citep{takahashi1985}.
Further details on the ecosystem model can be found in
\citet{maier2005,schwinger2016}.
We continue with a description of what is most relevant for the sediment, namely
the particles in the seawater.


Lithogenic clay is added into the top layer of the ocean, as a refractory tracer
of unknown composition (assuming quartz for its density) as well as a direct
input of iron through the instant dissolution of the deposited dust.
Besides dust input, there is river input of dissolved components into the ocean,
as well as of \chem{POC} \citep{tjiputra2020}.
We use a parameterisation for organic carbon particles that linearly increase
their sinking speed with depth to account, in a crude way, for aggregation.
The other particles, clay, \chem{CaCO_3} and \chem{bSiO_2}, sink through the
water column with a homogeneous velocity.
There is only one size class of each of the particles and each has a
unific reactivity.
The remineralisation rates for \chem{POC} and \chem{bSiO_2} are constant, whereas the
dissolution of \chem{CaCO_3} depends on its saturation state (using first-order
dissolution kinetics).

\subsubsection{Sediment processes}      \label{sec:sediment}
The model's sediment module contains four
solid compounds and seven solutes (Table~\ref{tab:tracers}).
Each of these tracers has a counterpart in the seawater.

\begin{table}[h]
\begin{tabular}{lll}
\toprule
symbol          & meaning               & units \\
\midrule
\multicolumn{3}{l}{\textit{state variables of the solid fraction ($s$)}} \\
\chem{OC}       & organic carbon        & \si{\mol\phosphorus\per\cubic\deci\metre} \\
\chem{CaCO_3}   & calcium carbonate     & \si{\mol\carbon\per\cubic\deci\metre} \\
\chem{bSiO_2}   & biogenic silica       & \si{\mol\silicon\per\cubic\deci\metre} \\
\chem{clay}     & lithogenic (from dust) & \si{\kilo\gram\per\cubic\metre} \\
\midrule
\multicolumn{3}{l}{\textit{state variables of the porewater solutes ($d$)}} \\
\chem{DIC}      & dissolved inorganic carbon & \si{\mol\carbon\per\cubic\deci\metre} \\
$A_T$           & total alkalinity      & \si{\equiv\per\cubic\deci\metre} \\
\chem{PO_4}     & phosphate             & \si{\mol\phosphorus\per\cubic\deci\metre} \\
\chem{O_2}      & oxygen                & \si{\mol\oxygen\per\cubic\deci\metre} \\
\chem{N_2}      & molecular nitrogen    & \si{\mole\nitrogen\per\cubic\deci\metre} \\
\chem{NO_3}     & nitrate               & \si{\mol\nitrogen\per\cubic\deci\metre} \\
\chem{Si(OH)_4} & silicic acid          & \si{\mol\silicon\per\cubic\deci\metre} \\
\bottomrule
\end{tabular}
\caption{Biogeochemical tracers in the sediment model.}
\label{tab:tracers}
\end{table}

The three biogenic particles---organic carbon, \chem{CaCO_3} and
\chem{bSiO_2}---and lithogenic clay become part of the sediment by sinking from
the bottom seawater layer into the upper layer of the sediment.
In reality, sedimentation (also called ``rain'' or ``deposition'') creates new
sediment on top of existing sediment, but in the model the sediment is defined
on a fixed grid.
The layer interfaces of the sediment are defined at 0, 0.1, 0.4, 0.9, 1.6, 2.5, 3.6, 4.9, 6.4,
8.1, 10.0, 12.1, 14.4 and 16.9\;cm below the seawater--sediment interface.
These define the twelve active layers, and one burial layer that stores all the
particulate tracers that have been transported all the way down out of the
active sediment domain.
Typically, most of the sedimented material remineralises within the sediment;
only a fraction of the biogenic compounds is buried.
In the model, clay is considered inert and does not dissolve.
In a steady state, burial equals sedimentation minus dissolution.

For each biogenic solid compound $s$ there is at least one dissolved porewater
component $d$.
Within the sediment, these evolution equations hold:
\begin{subequations}
\begin{align}
\frac{\mathrm{d}c_s}{\mathrm{d}t} &= \mathcal{B} \frac{\partial^2 c_s}{\partial z^2}
                                - \frac{\partial}{\partial z}(w_s c_s)
                                - \frac{v_{s\rightarrow d} M_s}{\rho(1-\phi)} \label{eqn:solid} \\
\frac{\mathrm{d}c_d}{\mathrm{d}t} &= \frac{\partial}{\partial z}\left(\mathcal{D} \frac{\partial c_d}{\partial z}\right)
                                + \frac{v_{s\rightarrow d}}{\phi} \label{eqn:dissolved} \,,
\end{align}
\label{eqn:both}
\end{subequations}
where $c_s$ is the weight fraction of compound $s$ and $c_d$ the porewater
concentration of solute $d$, $\mathcal{B}$ is the diffusion
coefficient for bioturbation, $w_s$ the vertical advection velocity of the solid
compounds, $v$ the reaction rate, $M_s$ the molecular weight of the solid
compound, $\rho$ the bulk sediment density, $\phi$ is the porosity (the
volume of the porewater divided by the total volume including the target solid
sediment) and $\mathcal{D}$ is the diffusion coefficient for porewater
diffusion.

Sediment is displaced through bioturbation (Eq.~\ref{eqn:solid}, RHS, 1st term)
and vertical advection (2nd term) of solid constituents.
Each of the last terms of Eqs~\ref{eqn:solid} and~\ref{eqn:dissolved} denotes
decomposition of the solid components ($s\rightarrow d$ conversion).

Organic matter, usually referred to as ``organic carbon'', consists of many
organic compounds of which the major ones are modelled.
When organic matter remineralises, we assume that all those compounds end up in
dissolved phase simultaneously.
In other words, in addition to the principal dissolved components $d$ (phosphate
for organic matter) from Eq.~\ref{eqn:dissolved} there are other products $d'$
(e.g.\ nitrate) that are created according to:
\begin{equation}
\frac{\mathrm{d}c_{d'}}{\mathrm{d}t} = R \, \frac{\mathrm{d}c_{d}}{\mathrm{d}t} \,,
\label{eqn:dd}
\end{equation}
where $R$ is a stoichiometric expression, relating $d'$ to $d$.
Similarly, total alkalinity decreases as a consequence of organic carbon
remineralisation and increases with the dissolution of calcium carbonate.


The porosity $\phi$ is a prescribed function of depth, starting at 0.85 in the
upper layer decreasing to 0.62 in the twelfth layer at around 13\;cm depth.
The rest of the volume, $1-\phi$, is thus 0.15 at the surface and increases,
through compactification, to 0.38 at 13\;cm depth.
In a filled state, this is the volume fraction of the bulk solid matter.

The model simulates the advection with respect to the sediment--water
interface by shifting sediment vertically, depending on the ``filling'' state of
the sediment.
The total volume fraction occupied by solid material is given by the sum of the
components' volume fractions:
\begin{equation}
V_\mathrm{f} = \sum_s \frac{M_s}{\rho_s} \, c_s   \,,
\end{equation}
where $s$ runs over the four types of particulate matter, $\rho_s$ is the
density of the solid component (thus $M_s/\rho_s$ the specific volume)
and $c_s$ the concentration.
The shifting procedure sees to it that $V_\mathrm{f}$ approaches $1-\phi$.
If there is more deposition than remineralisation, sediment moves downwards
at all depths to fill up the layers and the surplus created in the bottom layer
of the active sediment (layer 12) is moved into the burial layer (layer~13).
If there is not enough deposition relative to remineralisation, sediment moves
upwards.
To this end, the active sediment's bottom layer will be overfilled from the
burial layer with enough clay to fill the whole sediment.
Then the layers above will be filled by redistribution (upward advection) from the
surplus in layer~12.

\subsubsection{Sediment--seawater burst coupling}   \label{sec:burst}

\begin{figure}
    \centering
    \includegraphics[width=.72\linewidth]{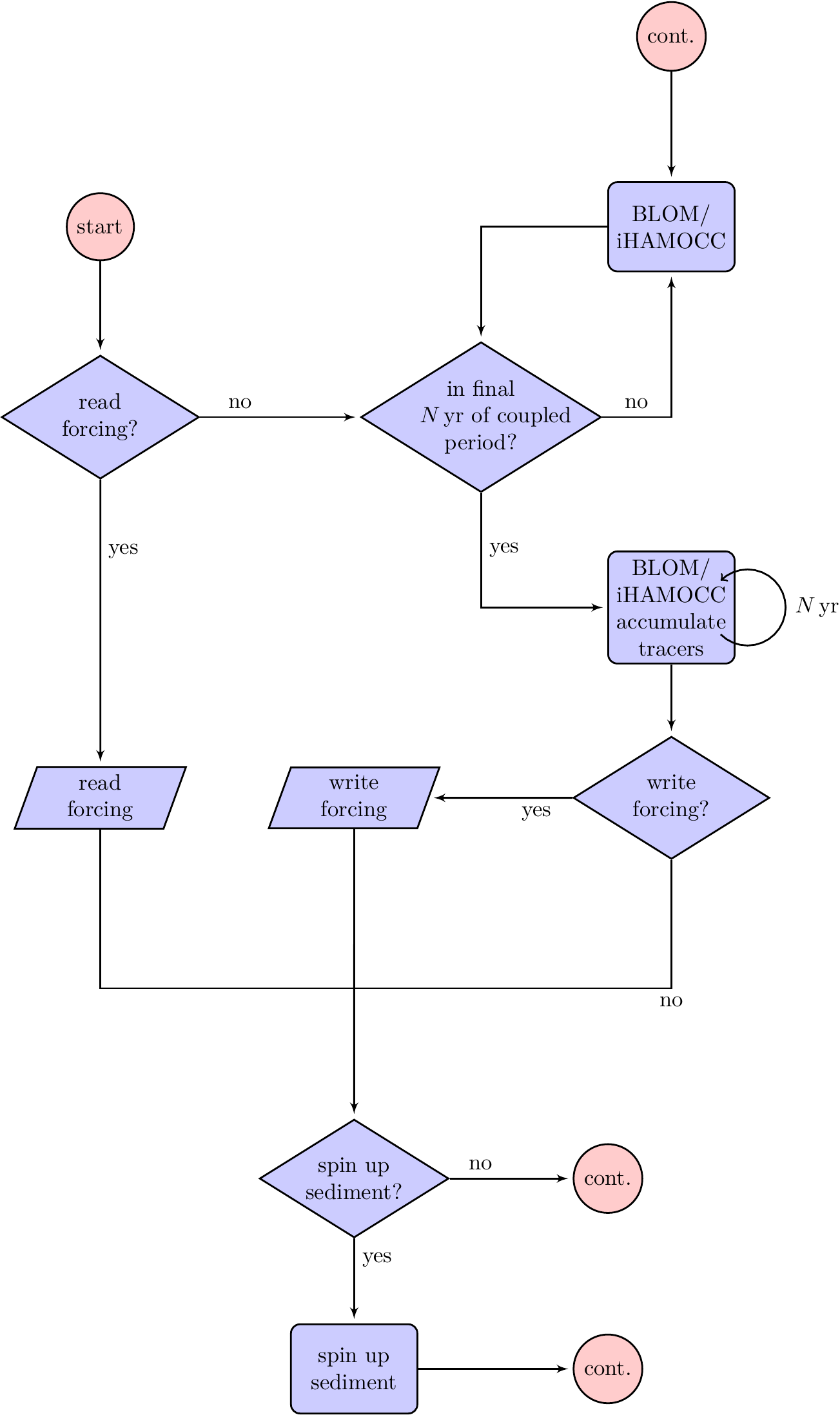}
    \caption{Flowchart of the burst coupling.
    The circles are connectors, the rectangles are processes and the rhomboids
    present input/output processing.}
    \label{fig:flowchart}
\end{figure}

In order to simulate a near-equilibrium of sediment biogeochemistry that is in
agreement with the ocean circulation and biogeochemistry, we developed
the so-called ``burst-coupling'' method.
The model, BLOM/iHAMOCC, is adjusted such that we can choose to simulate the
sediment processes alone while being forced by a seasonal climatology of
bottom-seawater variables, or to couple the sediment with the rest of the ocean
model.
The variables include particle fluxes from the ocean onto the sediment and
tracer concentrations.
The fluxes define the sedimentation of biogenic and terrigenic particles.
Dissolved tracer concentrations are used to calculate gradients between the
porewater and overlying seawater for sediment--seawater fluxes.
The climatology of the bottom-seawater variables consists of \SI{12}{\month}
and may be an average over a certain number of years at the end of the preceding
coupled period.

Normally, a single iteration step for the sediment (finite difference
approximation of the differential equations of Sec.~\ref{sec:sediment}) is
performed for each water column biogeochemistry iteration, but in stand-alone
mode a huge many iterations of the sediment are done (with a much larger
timestep to speed up even more).
These modes can be alternated in an automated manner where the user defines the
coupled-mode and the decoupled-mode periods.
Figure~\ref{fig:flowchart} shows the flowchart for this process.

If there is a bottom-seawater forcing present, it can be read from file.
Otherwise, the simulation starts in coupled mode and a user-defined number of
years are accumulated during all or the last part of the simulation (at least
the last year) to derive a 12-month climatology.
Optionally, this climatology is copied from volatile memory to file.
Then the sediment is spun up decoupled.
The process may be iterated by continuing the coupled simulation (with the
spun-up sediment) and so on.
The model terminates within the coupled mode when it reaches the user-defined
number of coupled model years (standard NorESM configuration); this is not
presented in this flowchart.

\subsubsection{Other improvements}
Much of the code has been refactored and tidied up, which improved readability
and computational performance (about 20\,\%).
The details can be found in the commits, see
\url{https://hg.gnu.org.ua/hgweb/burst-coupling/} and related discussion in
Appendix~\ref{app:recommendations}.

\subsection{Simulations}

\begin{figure}
    \centering
    \includegraphics[width=\linewidth]{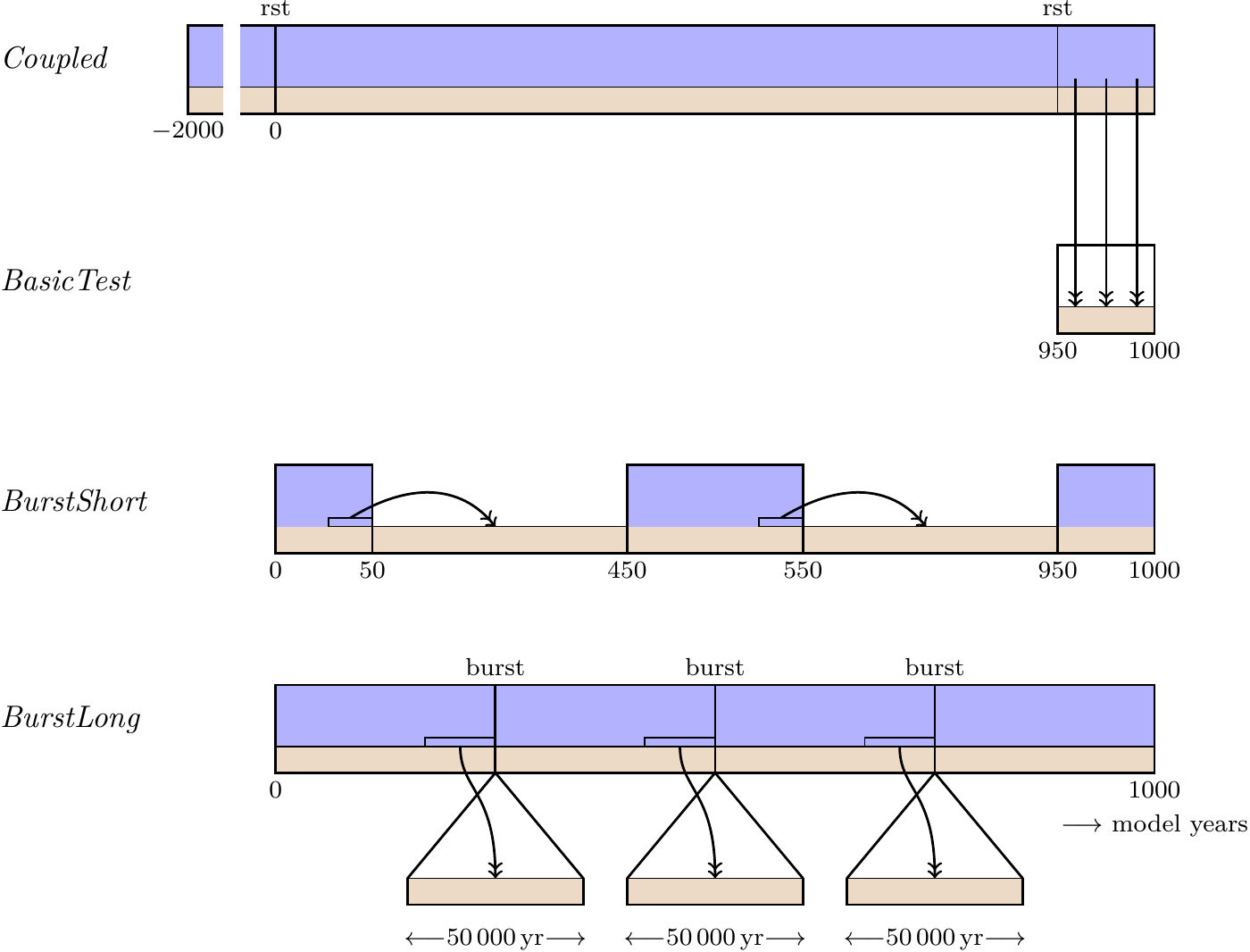}
    \caption{The model-time evolution of the four simulations (neither depth nor
    time are to scale).
    The blue areas represent the seawater and the brown the area below
    the seafloor (sediment and porewaters).
    The small rectangles in the lower-right corner of the seawater in the burst-coupling
    simulations represent the yearly averaged bottom-water climatology.
    The double arrows represent the forcing of off-line bottom-water variables
    onto the sediment.}
    \label{fig:simulations}
\end{figure}

We use a Holocene (pre-industrial) climate state.
We first run the ocean model for 2000\;yr, which is when the seawater tracers
reach an approximate steady state (but still drift because of the unfilled
sediment).
Then we use the final state of this simulation as the initial conditions for the
three different simulations that are listed in Table~\ref{tab:simulations} and
explained below.
The time evolution of each of the simulations is presented in
Fig.~\ref{fig:simulations}.

\begin{table}[h]
\begin{tabular}{lll}
\toprule
Name        & Duration                          & Iterations   \\
\midrule
Coupled     & 1000\;yr coupled                  & 1 \\
BurstShort  & 50\;yr coupled, 400\;yr sediment, 50\;yr coupled & 2 \\
BurstLong   & 250\;yr coupled, 50\;kyr sediment & 4 \\
\bottomrule
\addlinespace
\end{tabular}
\caption{Numerical simulations performed.
All the simulation are forked from a 2000\;yr coupled simulation, such that the
water column is in an approximate steady state to start with.}
\label{tab:simulations}
\end{table}

\begin{figure}
    \centering
    \includegraphics[width=\linewidth]{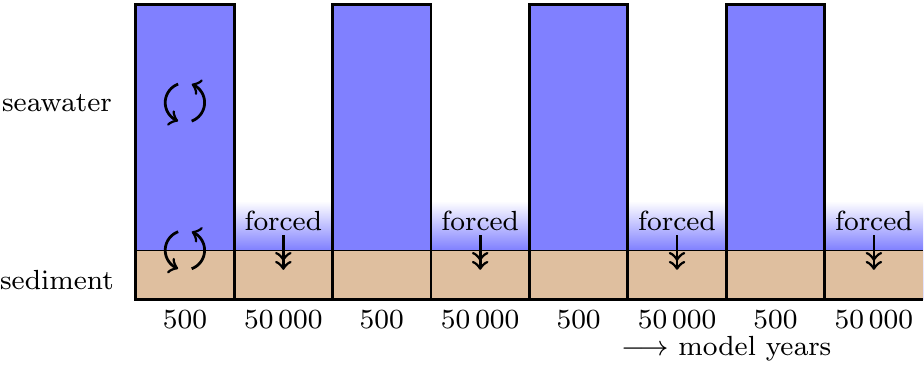}
    \caption{Burst coupling, a method to efficiently spin up a model consisting
    of a computationally slow and a fast component, where, in order to reach a
    steady state, the fast component needs to run many more model years than the
    slow component.
    In the depicted example, the coupled BLOM/iHAMOCC model that includes the
    computationally slow seawater component, is spun up for 250\;yr, after which
    the fast sediment component is spun up for a much longer time.
    This may be repeated a couple of times, each time with a recent forcing.}
    \label{fig:burst}
\end{figure}

\begin{itemize}
\item
\emph{Coupled} is a 1000-yr simulation where the water column and the
sediment are coupled.
Its purpose is that the other simulations can be compared to a base.
Like the other simulations, it is preceded by a 2000-yr initial spin-up; so at
the end of \emph{Coupled}, the model has done 3000\;yr of coupled integration.
\item
\emph{BurstShort} has the same duration, but two intervals of it are sediment-only,
decoupled from the water column and forced by bottom water particle fluxes and
tracer concentrations calculated from the last 50\;yr of the preceding coupled
simulation interval.
This simulation tests if the method is basically working and if the
sediment---when forced by a bottom-water climatology---is evolving sufficiently
similar to \emph{Coupled}.
In other words, this shows how the water column forcing (in contrast to coupled
mode) is affecting the sediment.

\item
\emph{BurstLong} is a simulation of a total model duration of 151\,000\;yr.
During the last 50\;yr of an initial 250\;yr coupled spin-up, the
flux and tracer values are averaged over the duration each month and over the
50\;yr.
Then we run only the sediment model, decoupled from the rest of the model for
50\,000\;yr and forced by this climatology.
This is iterated three times and finalised with another 250\;yr of coupled
integration (Fig.~\ref{fig:burst}).
By interleaving every 250\;yr of coupled simulation with 50\;kyr of
sediment-only, we manage to make the total resulting coupled simulation
1000\;yr.
Hence we can analyse the effect of the long sediment bursts to the water column
(compared to the 1000\;yr of \emph{Coupled}).
Additionally, the simulation may give insight in how long an ocean model needs
to spin up (without drifting because of sediment interaction).
\end{itemize}

\subsection{Observations and evaluation}

We use data from \citet{archer1996,data::archer1999} to evaluate the \chem{CaCO_3} and
\chem{bSiO_2} mass fractions in the sediment.
Furthermore, the burial fluxes from \citet{hayes2021} are used as a further
criterion for the model performance.

\section{Results}                       \label{sec:results}

\subsection{Coupled simulation}

\begin{figure}[!hb]
    \centering
    \includegraphics[width=\linewidth]{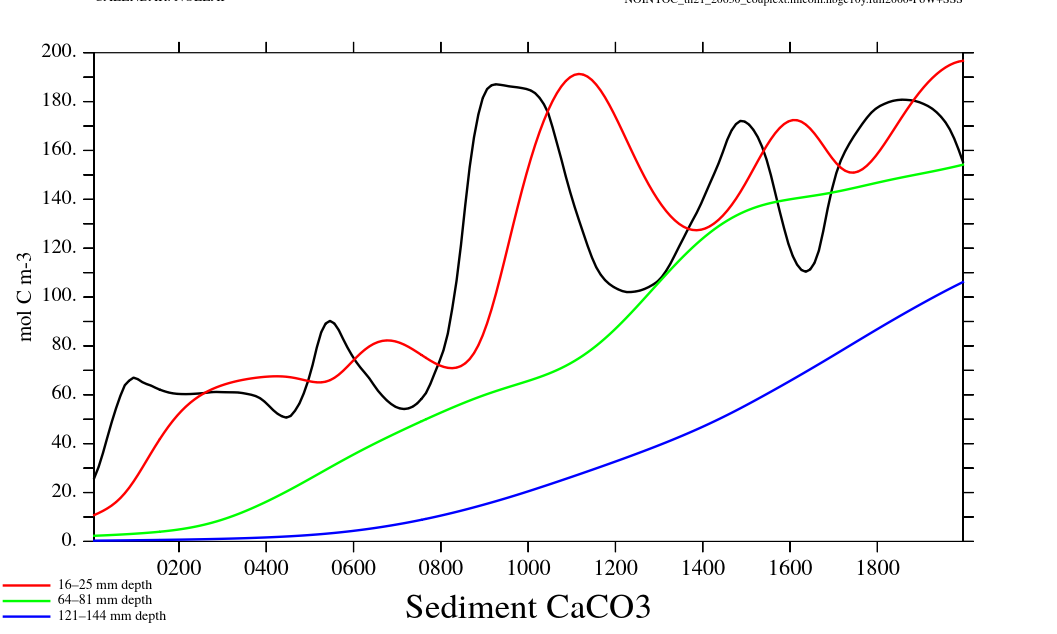}
    \caption{\chem{CaCO_3} timeseries at different depths in the sediment of
    \emph{Coupled} in the Gulf of Panama (80\degree\,W, 5\degree\,N).}
    \label{fig:ts:calcite:Coupled:Panama}
\end{figure}

\begin{figure}
    \centering
    \includegraphics[width=\linewidth]{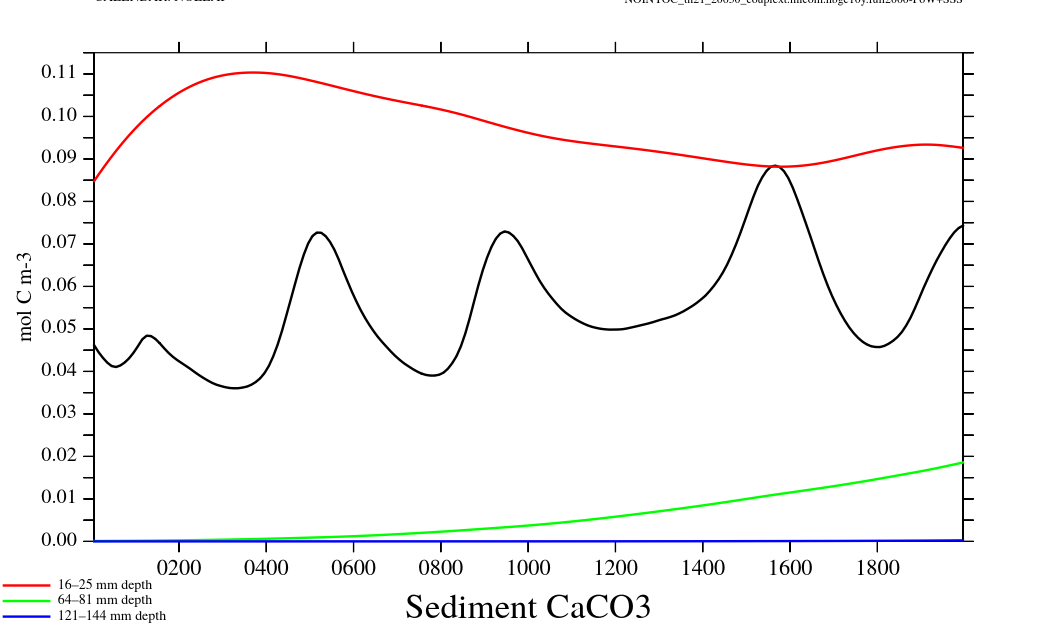}
    \caption{\chem{CaCO_3} timeseries at different depths in the sediment of
    \emph{Coupled} at MANOP site~S (140\degree\,W, 11.33\degree\,N).}
    \label{fig:ts:calcite:Coupled:MANOP}
\end{figure}

\begin{figure}
    \centering
    \includegraphics[width=\linewidth]{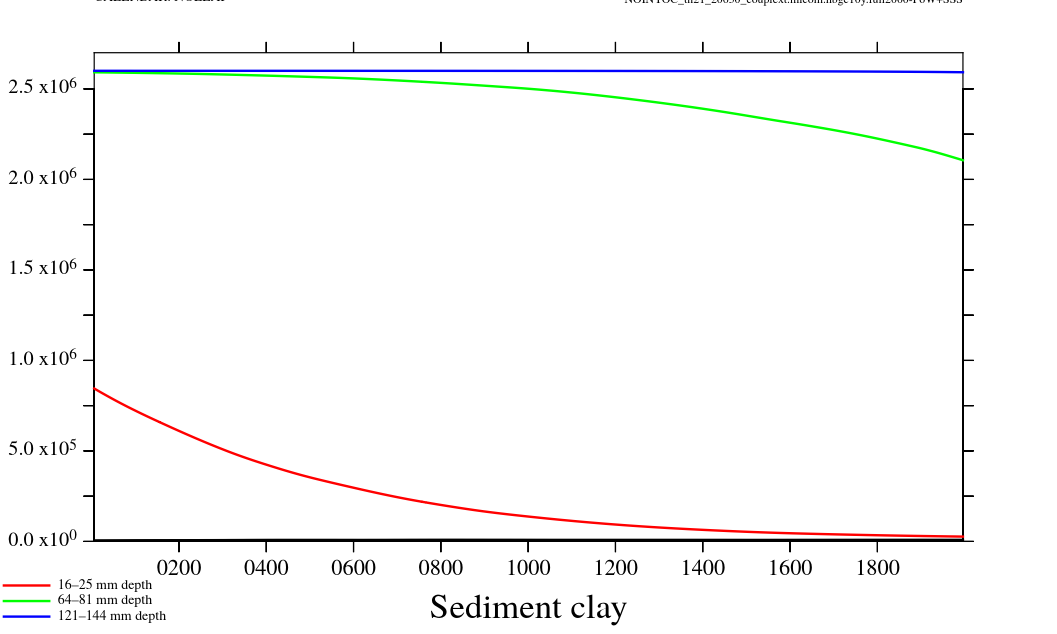}
    \caption{Clay timeseries at different depths in the sediment of
    \emph{Coupled} at MANOP site~S (140\degree\,W, 11.33\degree\,N).}
    \label{fig:ts:clay:Coupled}
\end{figure}

\begin{figure}
    \centering
    \includegraphics[width=\linewidth]{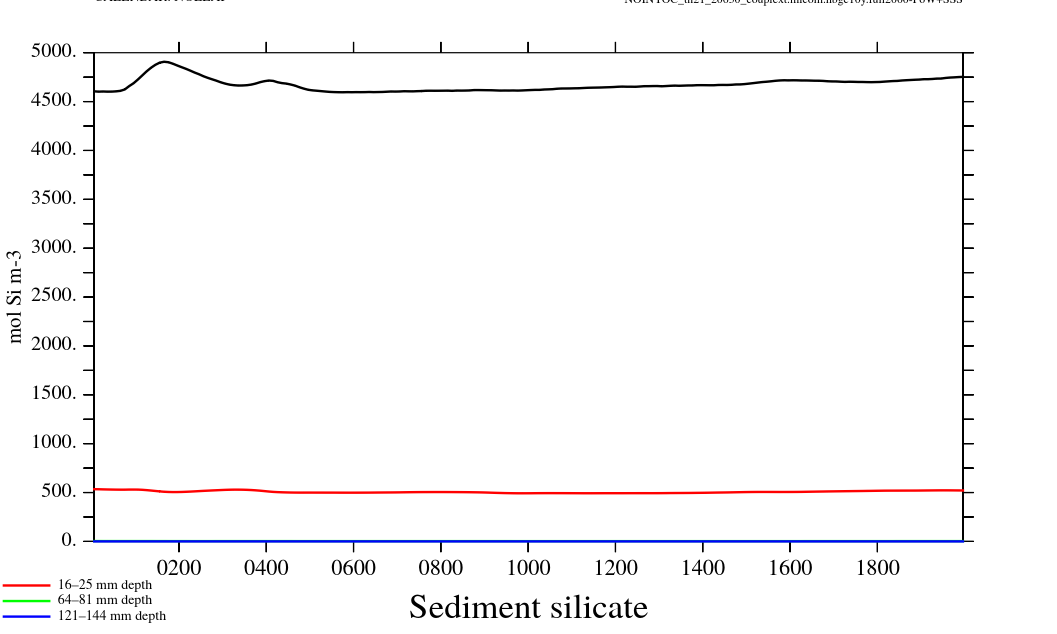}
    \caption{b\chem{SiO_2} timeseries at different depths in the sediment of
    \emph{Coupled} in the Gulf of Panama (80\degree\,W, 5\degree\,N).}
    \label{fig:ts:bSi:Coupled}
\end{figure}

In Figs~\ref{fig:ts:calcite:Coupled:Panama}--\ref{fig:ts:bSi:Coupled} we
present porewater and particle concentration
development over the 2000 or 3000\;yr spin-up of the coupled simulation.
The figures show that the sediment is not filled with (most) particles within a
couple of thousand years.
For instance, in the Gulf of Panama \chem{CaCO_3} appears to reach a dynamical
equilibrium after about 1000\;yr in the upper layers of the sediment (upper
2.5\;cm), but at higher depths (below 6.4\;cm) it will take notably longer than
2000\;yr for \chem{CaCO_3} to reach a steady state
(Fig.~\ref{fig:ts:calcite:Coupled:Panama}).
Only b\chem{SiO_2} looks in a steady state from the start, or maybe from about
500\;yr.

This alone warrants a relatively long spin-up of at least the sediment.
Even if one is not interested in the sediment, it should be realised that the
interaction of the seawater with a changing (non-steady state) sediment will
lead to a drift in an ocean- or climate model.
Of course a priori reasoning is not enough to conclude if such a drift is
significant.

%
%

\subsection{Global spin-up evolution}
\begin{figure}[!hb]
    \centering
    \includegraphics[width=\linewidth]{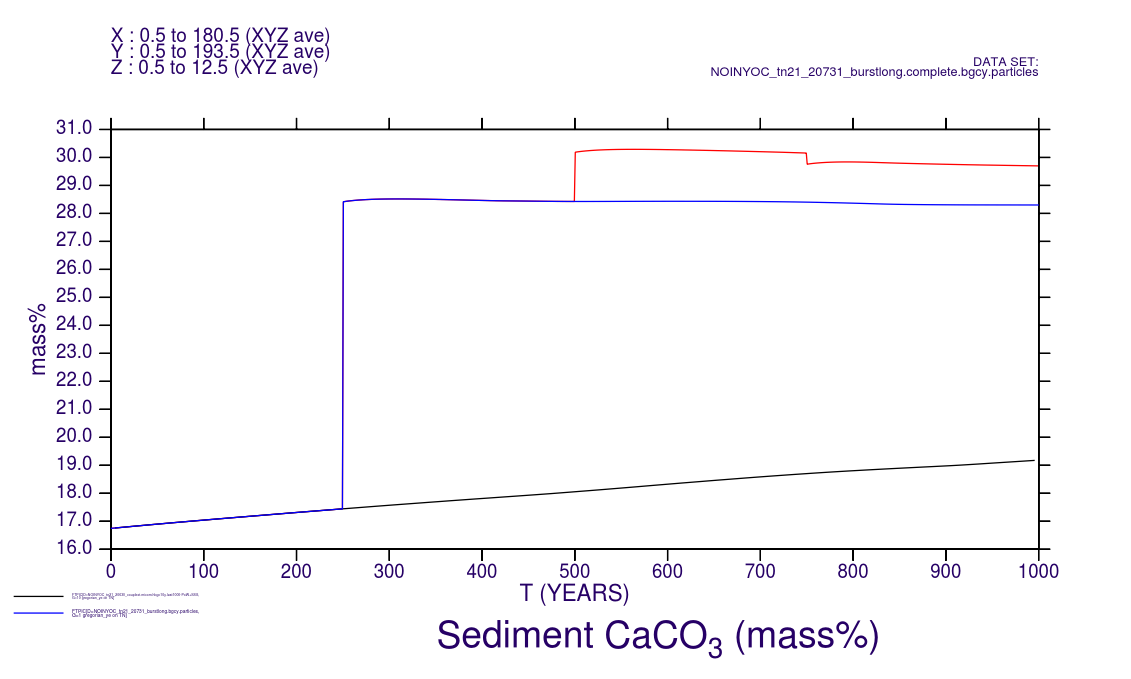}
    \caption{Modelled \chem{CaCO_3} timeseries in the sediment (global average,
    upper 14.4\;cm): black is 1000\;yr of coupled, blue includes one
    sediment-only period of 50\,000\;yr, red includes three such burst periods.}
    \label{fig:ts:calcite-alt}
\end{figure}
\begin{figure}
    \centering
    \includegraphics[width=\linewidth]{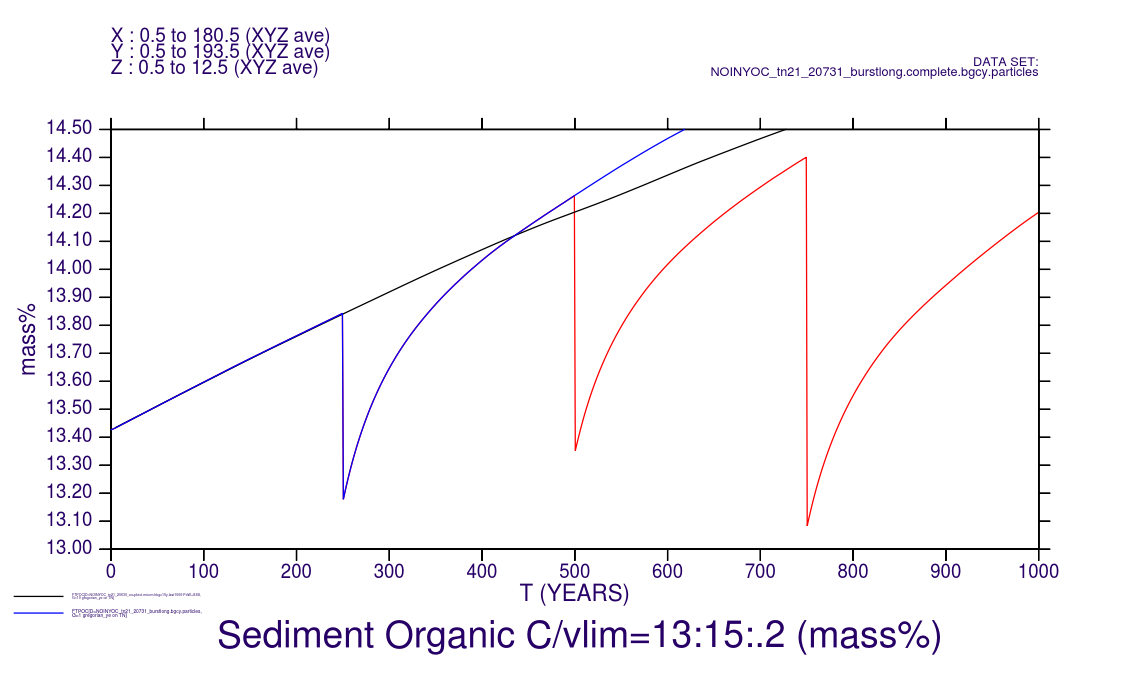}
    \caption{Modelled \chem{OC} timeseries in the sediment (global average,
    upper 14.4\;cm): black is 1000\;yr of coupled, blue includes one
    sediment-only period of 50\,000\;yr, red includes three such burst periods.
    See Fig.~\ref{fig:ts:oc-inter} for the timeseries including the 50\,000\;yr
    decoupled bursts (albeit compressed).}
    \label{fig:ts:oc-alt}
\end{figure}
\begin{figure}
    \centering
    \includegraphics[width=\linewidth]{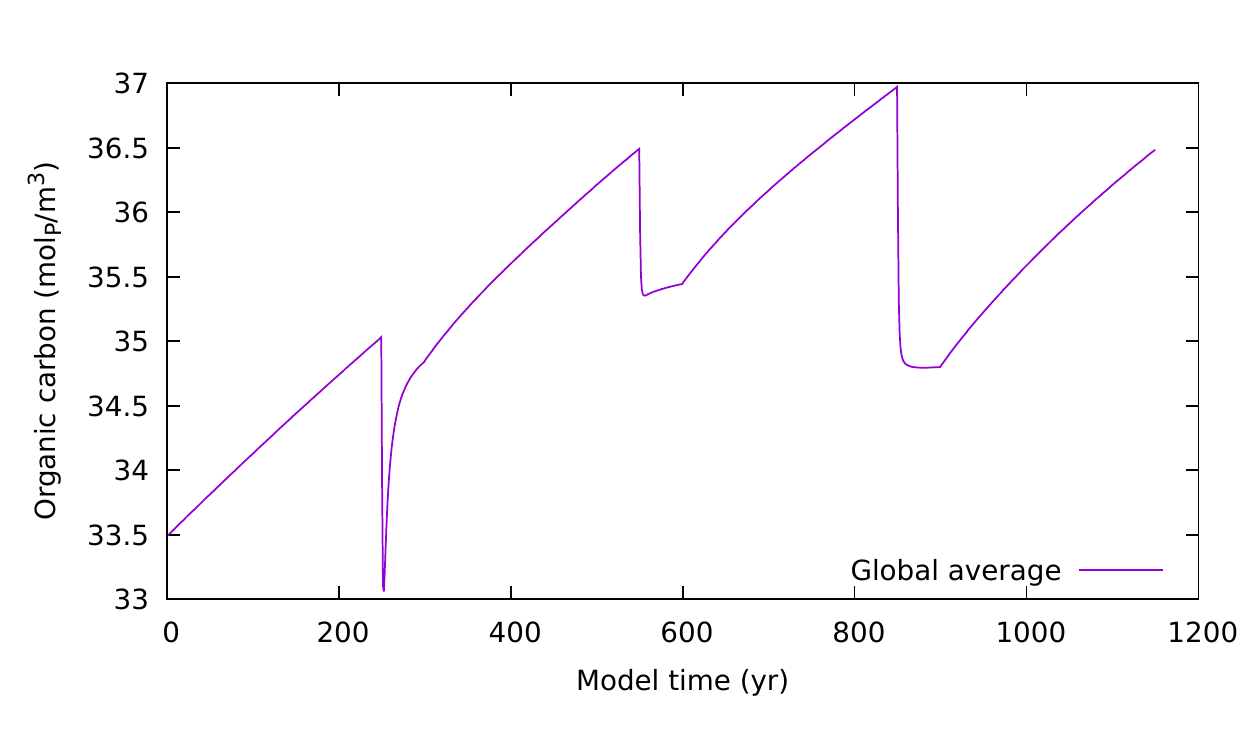}
    \caption{Modelled \chem{OC} timeseries in the sediment (global average,
    upper 14.4\;cm) from \emph{BurstLong}.  The time axis of the 50\,000\;yr
    burst periods are compressed by a factor of~1000.}
    \label{fig:ts:oc-inter}
\end{figure}
\begin{figure}
    \centering
    \includegraphics[width=\linewidth]{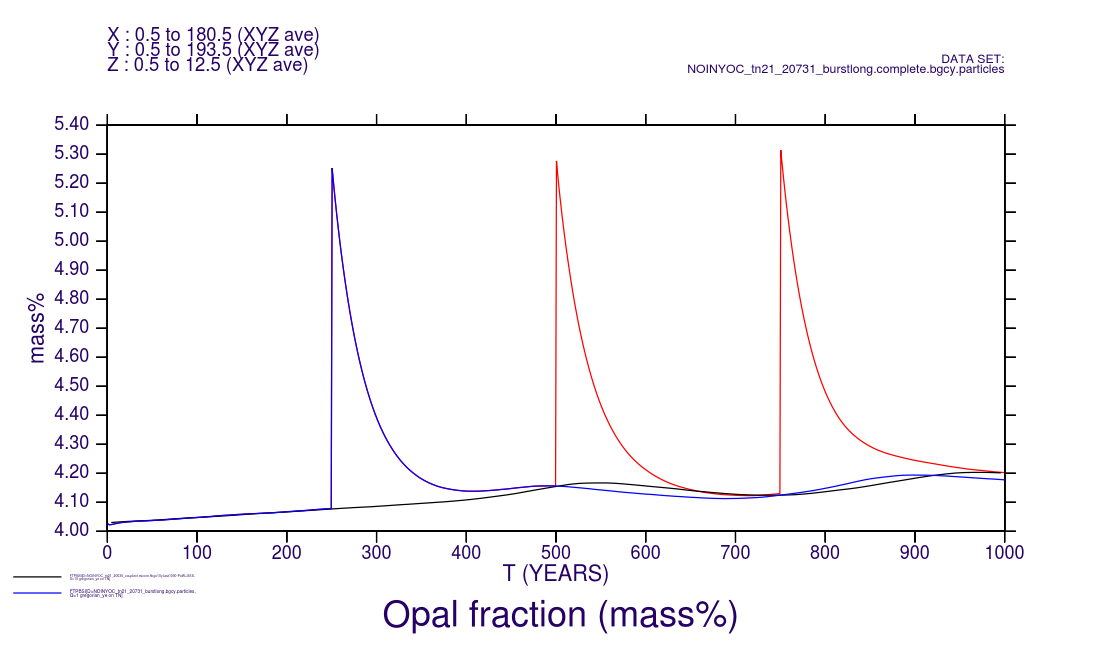}
    \caption{Modelled \chem{bSiO_2} timeseries in the sediment (global average,
    upper 14.4\;cm): black is 1000\;yr of coupled, blue includes one
    sediment-only period of 50\,000\;yr, red includes three such burst periods.}
    \label{fig:ts:opal-alt}
\end{figure}
\begin{figure}
    \centering
    \includegraphics[width=\linewidth]{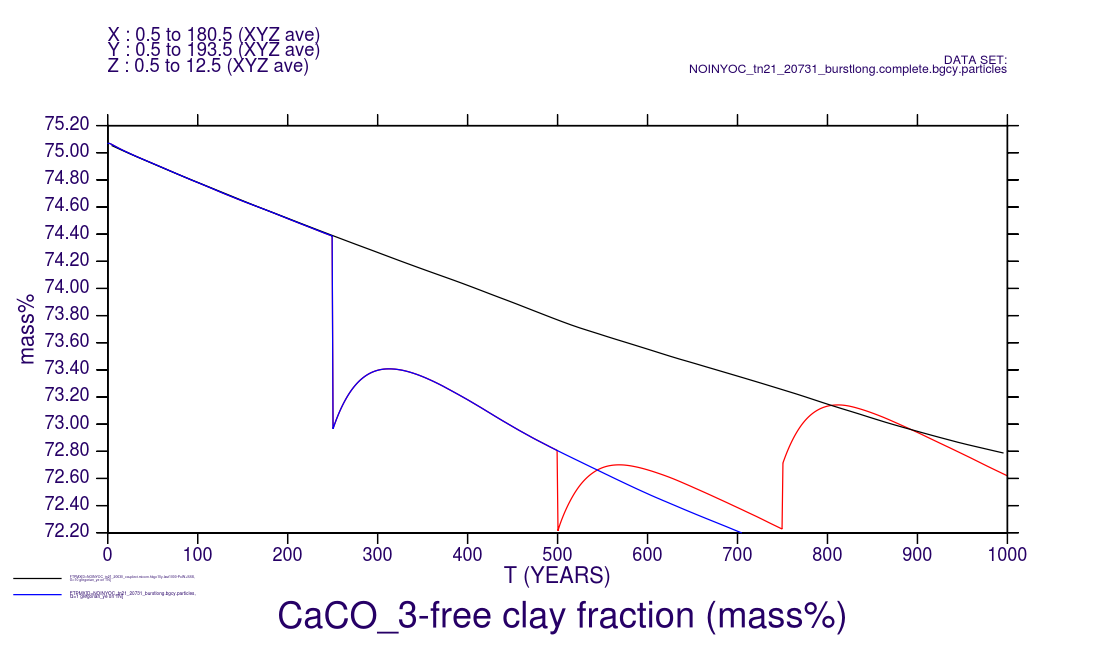}
    \caption{Modelled \chem{clay} timeseries in the sediment (global average,
    upper 14.4\;cm): black is 1000\;yr of coupled, blue includes one
    sediment-only period of 50\,000\;yr, red includes three such burst periods.}
    \label{fig:ts:clay-alt}
\end{figure}

We present the timeseries of \chem{CaCO_3} (sediment average) of the
\emph{BurstLong} simulation (Fig.~\ref{fig:ts:calcite-alt}).
It shows the 1000\;yr of coupled simulation, interleaved with the three
50\,000\;yr of decoupled sediment simulation.
The figure shows that at least two sediment-only burst
periods are needed to get the sediment in a steady state, and subsequent bursts
have less and less influence.
The bursts have a significant effect for the other particles as well, as can be
seen in Figs~\ref{fig:ts:oc-alt}--\ref{fig:ts:clay-alt}.
However, after each burst there is a relaxation towards the mass fractions of
the coupled simultion (\emph{Coupled}).
The relative change over the whole simulation time is much smaller for
organic carbon (Fig.~\ref{fig:ts:oc-alt}) and \chem{bSiO_2}
(Fig.~\ref{fig:ts:opal-alt}) than for \chem{CaCO_3}, and the deviation by the
bursts could just be contingent upon the forcing integrated.

\begin{figure}
    \centering
    \includegraphics[width=\linewidth]{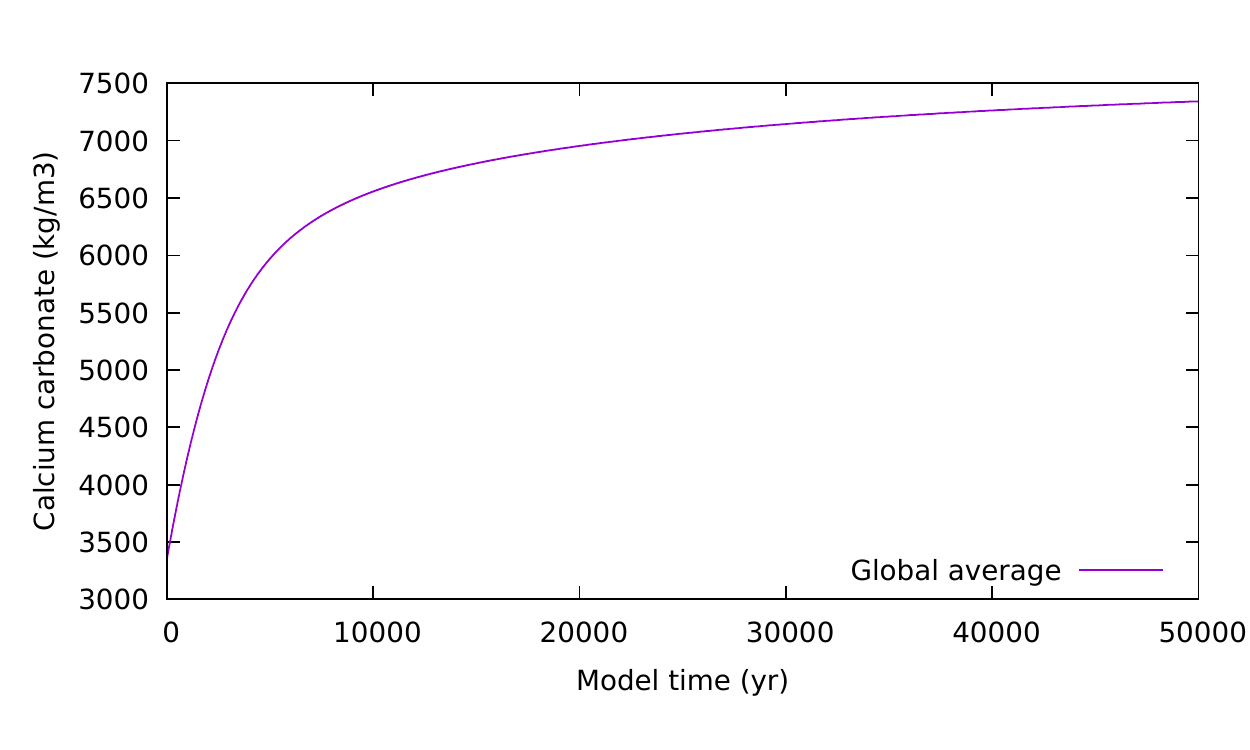}
    \caption{Modelled \chem{CaCO_3} timeseries in the sediment (average over
    upper 14.4\;cm) of the first `burst' of 50\,000\;yr.}
    \label{fig:ts:calcite-50kyr}
\end{figure}

\begin{figure}
    \centering
    \includegraphics[width=\linewidth]{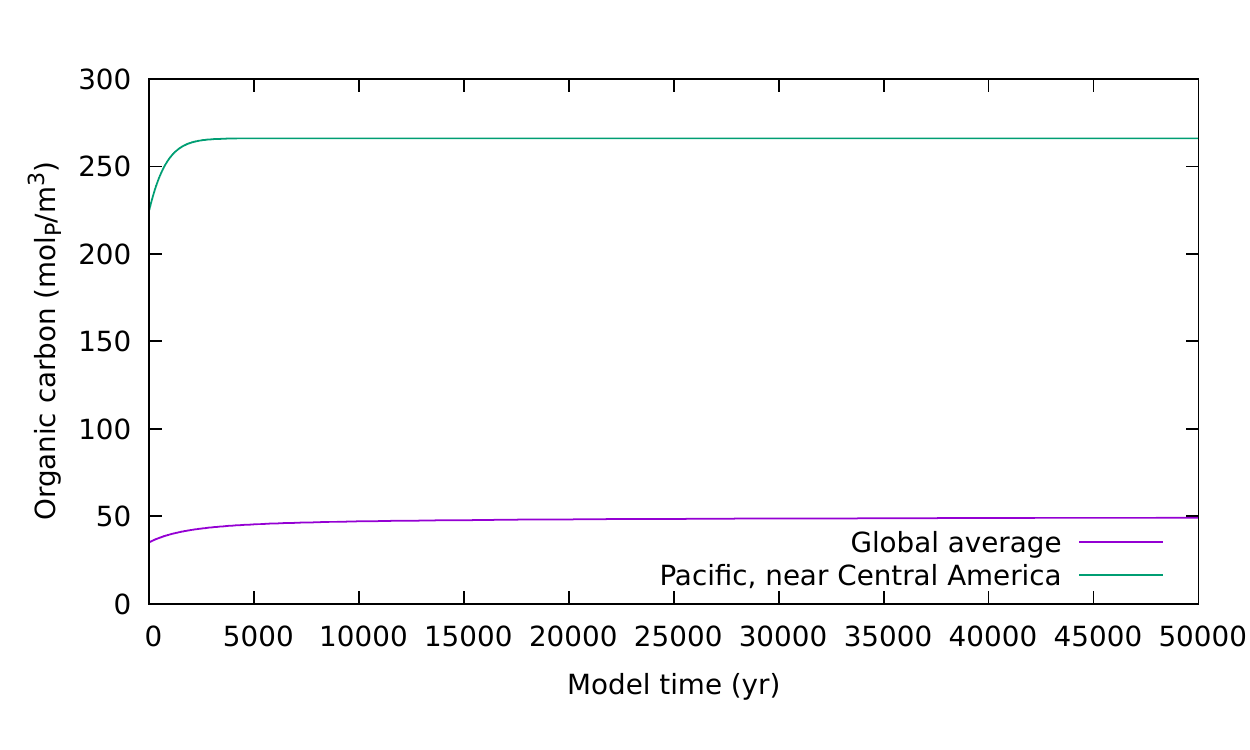}
    \caption{Modelled organic carbon (\si{\mole\phosphorus\per\cubic\metre}) timeseries in the sediment (average over
    upper 14.4\;cm) of the first `burst' of 50\,000\;yr.
    In addition to a global average the figure also shows the timeseries of the
    high particle flux region just west of Central America.}
    \label{fig:ts:oc-50kyr}
\end{figure}

Fig.~\ref{fig:ts:calcite-50kyr} shows a 50\;kyr burst, again from
\emph{BurstLong}, where the sediment content of \chem{CaCO_3} approaches a
steady state.
In the subsequent figure (Fig.~\ref{fig:ts:oc-50kyr}), organic carbon is
plotted for the same stand-alone sediment simulation.
In addition, also a region in the eastern Central Pacific is plotted where there
is a high flux of organic carbon (as well as biogenic silica).

\begin{figure}
    \centering
    \includegraphics[width=\linewidth]{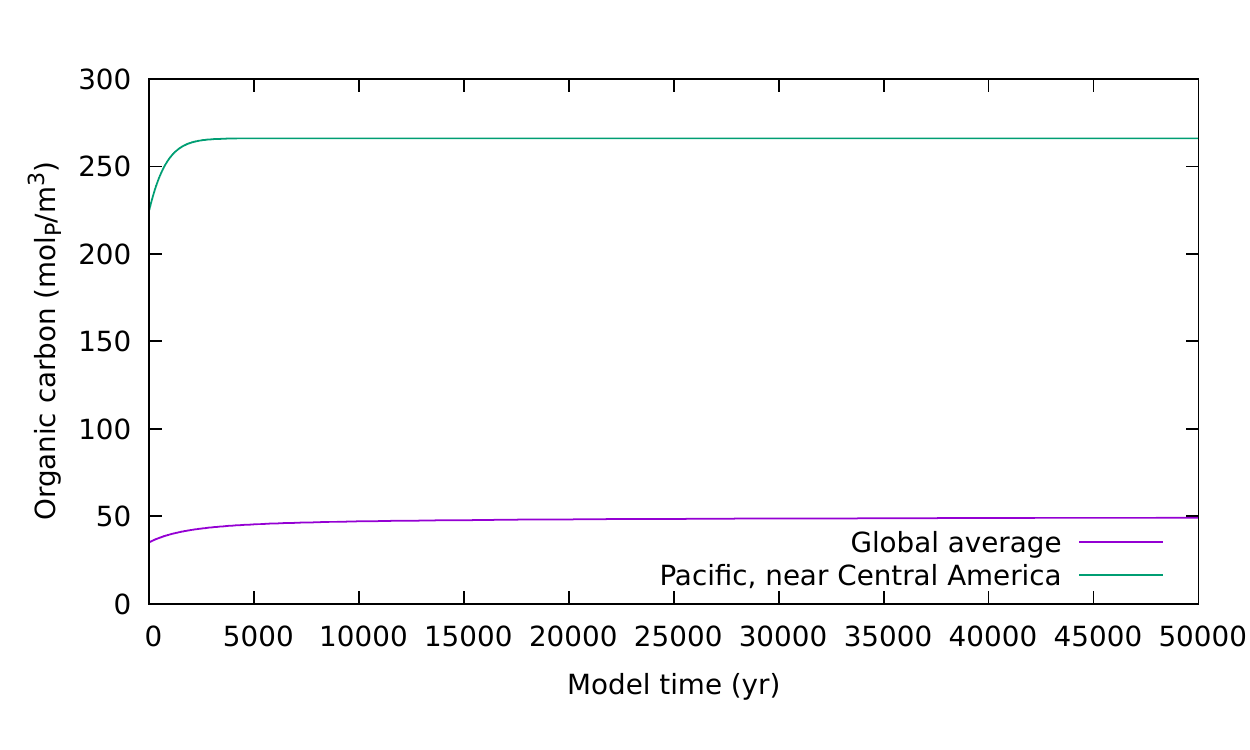}
    \caption{Modelled organic carbon (\si{\mole\phosphorus\per\cubic\metre})
    timeseries in the sediment (average over upper 14.4\;cm) of the full
    1000+150\,000\;yr simulation.
    The duration during the decoupled periods each 250\;yr is compressed such
    that one year on the x-axis represents 1000\;yr in model time.}
    \label{fig:ts:oc}
\end{figure}

As expected, during the 50-kyr sediment-only simulations immediately after years
250, 500 and 750 of the coupled simulation, much more burial occurs than during
any of the 250-yr periods.
We present the average molar concentration timeseries of organic matter in
Fig.~\ref{fig:ts:oc}.
Organic carbon changes only slightly during the 1000\;yr coupled + 150\;kyr
sediment-only simulation, from
\SIrange{33.5}{36.5}{\mole\phosphorus\per\cubic\metre}.
During coupled simulation, organic carbon increases (but the second derivative
is negative).
During sediment-only simulation, organic carbon decreases.

\subsubsection{Effect on water column}
\begin{figure}
    \centering
    \includegraphics[width=\linewidth]{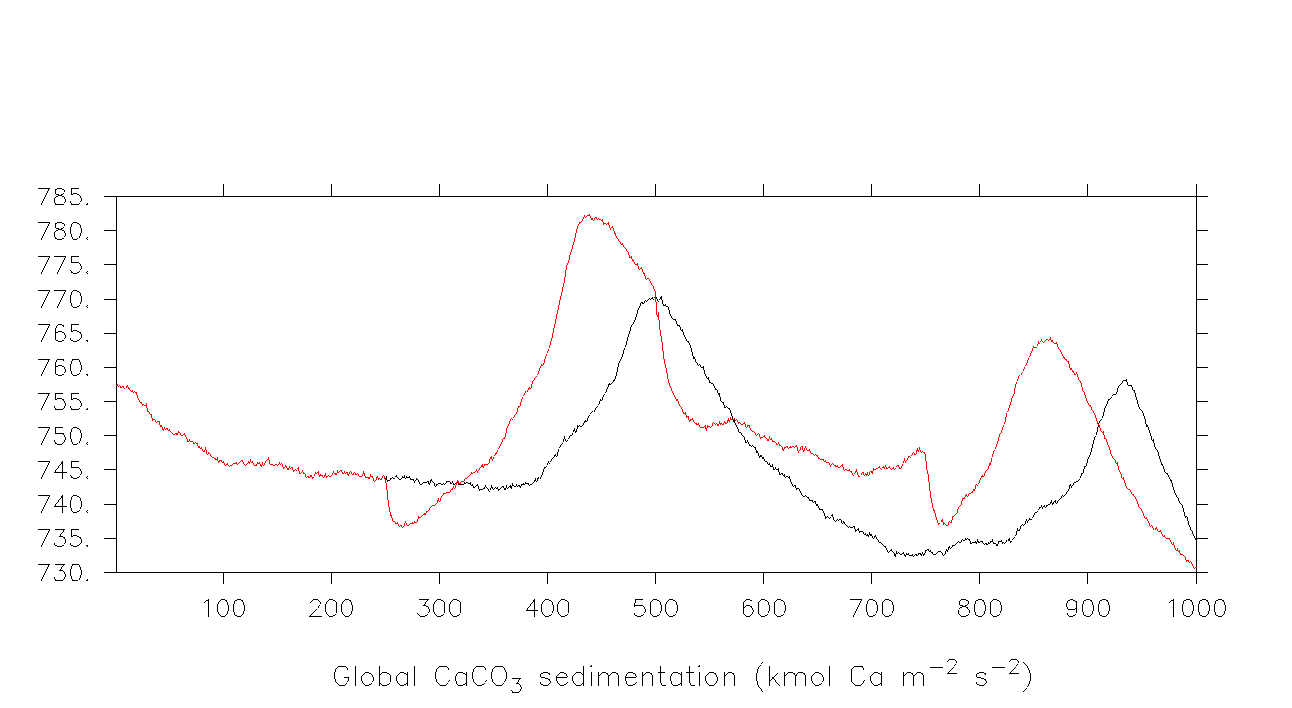}
    \caption{\chem{CaCO_3} global flux at the ocean bottom (sedimentation);
    against time (yr).
    The black line is the final 1000\;yr of \emph{Coupled}; the red line
    represents \emph{BurstLong}.}
    \label{fig:timeseries:jPiC_bot}
\end{figure}

\begin{figure}
    \centering
    \includegraphics[width=\linewidth]{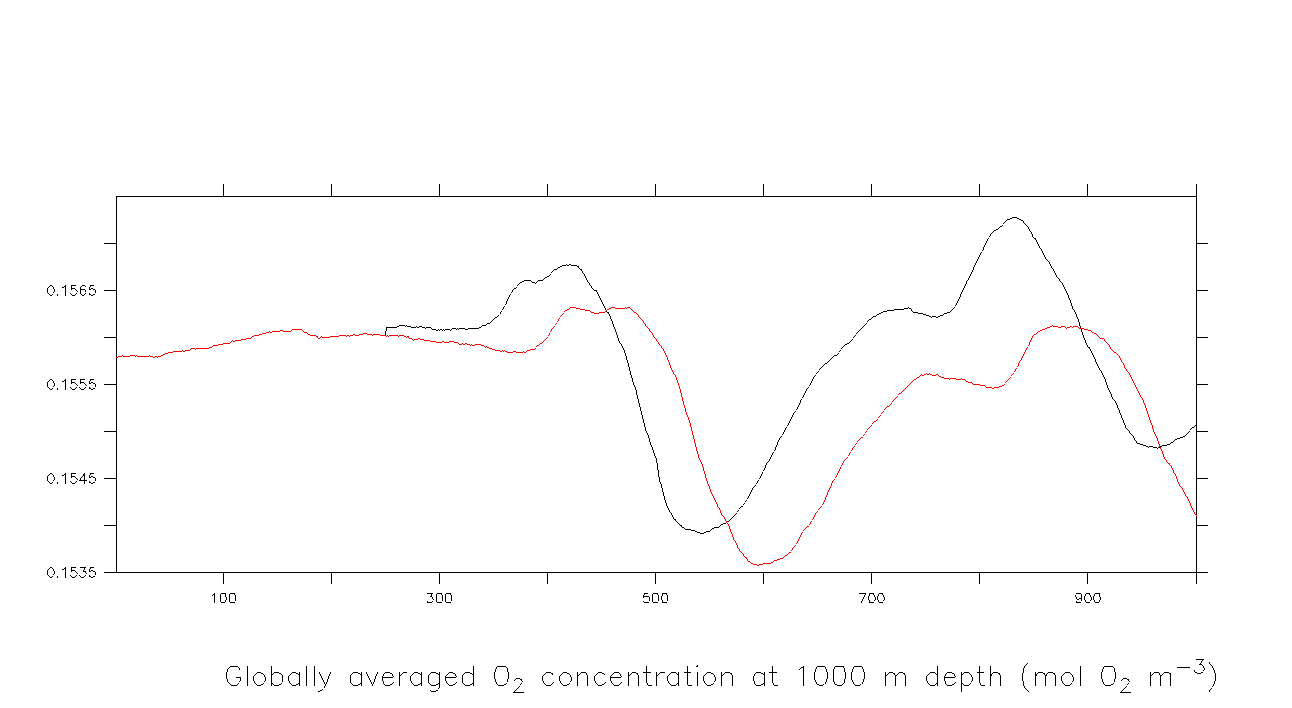}
    \caption{\chem{O_2} concentration at 1000\;m depth; global average;
    against time (yr).
    The red line is the final 1000\;yr of \emph{Coupled}; the black line
    represents \emph{BurstLong}.}
    \label{fig:timeseries:O2_1000m}
\end{figure}

\begin{figure}
    \centering
    \includegraphics[width=\linewidth]{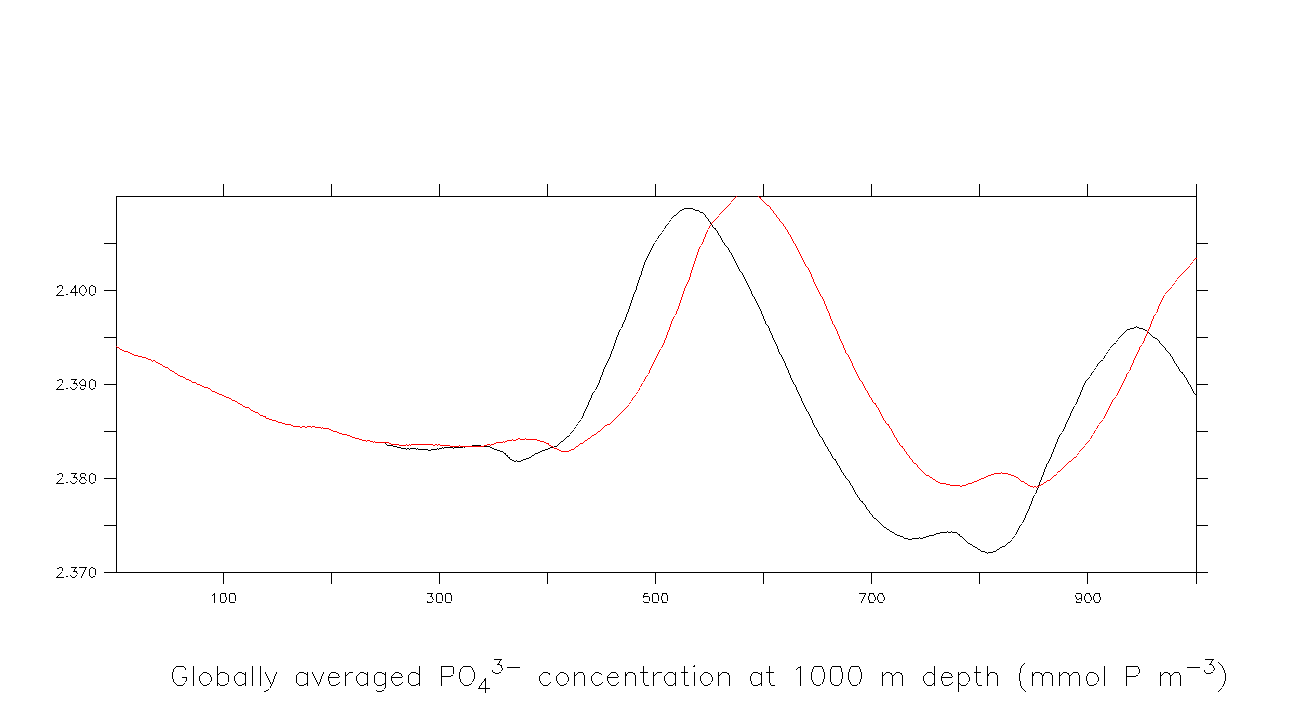}
    \caption{\chem{PO_4^{3-}} concentration at 3000\;m depth; global average;
    against time (yr).
    The red line is the final 1000\;yr of \emph{Coupled}; the black line
    represents the burst coupling with decoupled 50\;kyr periods every 250\;yr.}
    \label{fig:timeseries:PO4_1000m}
\end{figure}

Figs~\ref{fig:timeseries:jPiC_bot}--\ref{fig:timeseries:PO4_1000m} show
globally averaged quantities in the water column from \emph{Coupled} and
\emph{BurstLong}.
In all the figures from year 250, where the first decoupled period takes place,
the behaviour of \emph{BurstLong} (black line) is accelerated in both magnitude
and phase.
Of course the effect is much more moderate than for the tracers in the sediment,
but the 50\,000\;yr spin-up of the sediment clearly has an effect on the water
column tracers.
It is a small effect compared to the absolute value ($<1\%$) but large compared
to centenial variability ($>20\%$).

\subsection{Verification of the method}
\emph{BurstShort} would measure the effect of forcings on the sediment compared
to \emph{Coupled}.
We have not finished this analysis.
Instead we will show the results of a `synchronous' simulation, named
\emph{BasicTest}.
That simulation takes only 51\;yr of which 50\;yr forced by the respective
stored bottom water variables of \emph{Coupled} (see
Fig.~\ref{fig:simulations}).
As such we will, in a way, more precisely track the deviation from the coupled
simulation based on (about) the best presentation of bottom water forcings that
we can get while still using a full stored year.

\begin{figure}[!hb]
    \centering
    \includegraphics[width=\linewidth]{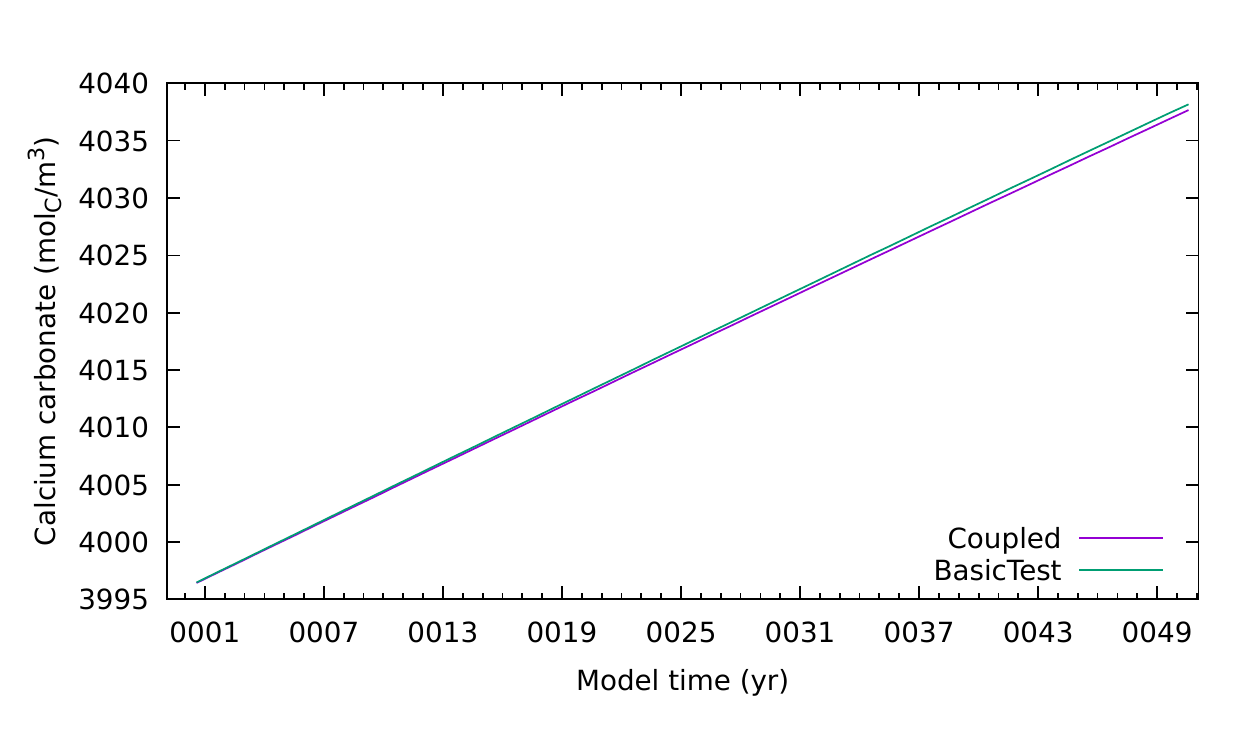}
    \caption{\chem{CaCO_3} molar concentration in the sediment of the last
    51\;yr of the simulations \emph{Coupled} and \emph{BasicTest}.}
    \label{fig:ts:calcite}
\end{figure}

\begin{figure}
    \centering
    \includegraphics[width=\linewidth]{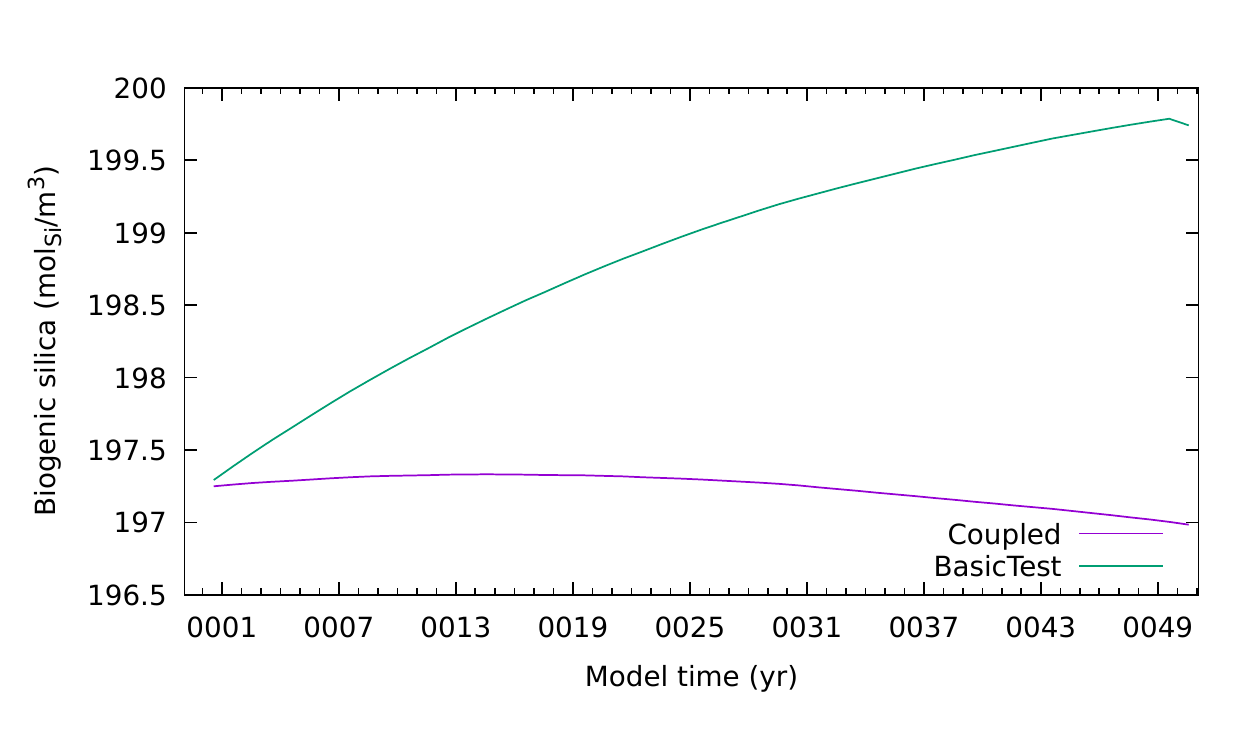}
    \caption{b\chem{SiO_2} molar concentration in the sediment of the last
    51\;yr of the simulations \emph{Coupled} and \emph{BasicTest}.}
    \label{fig:ts:opal}
\end{figure}

Figs~\ref{fig:ts:calcite} and~\ref{fig:ts:opal} present the timeseries of the
globally averaged sediment concetration of \chem{CaCO_3} and b\chem{SiO_2} of
\emph{Coupled} together with \emph{BasicTest}.
At the start of the simulations the concentration coincided for both tracers.
Then both tracers in \emph{BasicTest} start to deviate from \emph{Coupled}.
After 50\;yr, calcium carbonate has deviated about 0.01\,\%, whereas biogenic
silica has deviated about~1.4\,\%.

The water column concentrations did not change during the stand-alone sediment
simulation and continued spinning up only in the coupled periods (not further
presented).

\subsection{Evaluation}
This report will not show a full evaluation of the model that, amongst other
things, would compare the 151\,000\;yr simulation (\emph{BurstLong}) against
observations.
We have shown in the previous two subsections that our method works.
For a more complete evaluation of BLOM/iHAMOCC we refer to the analysis in the
recent coupled CMIP6 NorESM simulations \citep{tjiputra2020}.
One of the goals of our study was to show if a spun up sediment has an effect on
the ocean state and we have indications that it has.

\subsubsection{Pelagic zone}            \label{sec:pelagic}
BLOM/iHAMOCC has been evaluated by \citet{tjiputra2020}, but this is mostly
limited to the upper part of the water column.
Moreover, they had spun up the model in the order of 1000\;yr to a quasi steady
state.
As expected, most variables in the seawater from our \emph{BurstLong}
simulation are similar to those in \citet{tjiputra2020}.
The deepest part of the ocean and especially the sediment were not in a steady
state or not analysed.
Hence we will here evaluate a few features of the spun up model, mostly using
the average of the final part of \emph{BurstLong}.

\subsubsection{Benthic zone and seabed} \label{sec:benthic}
\begin{figure}[!hb]
    \centering
    \includegraphics[width=\linewidth]{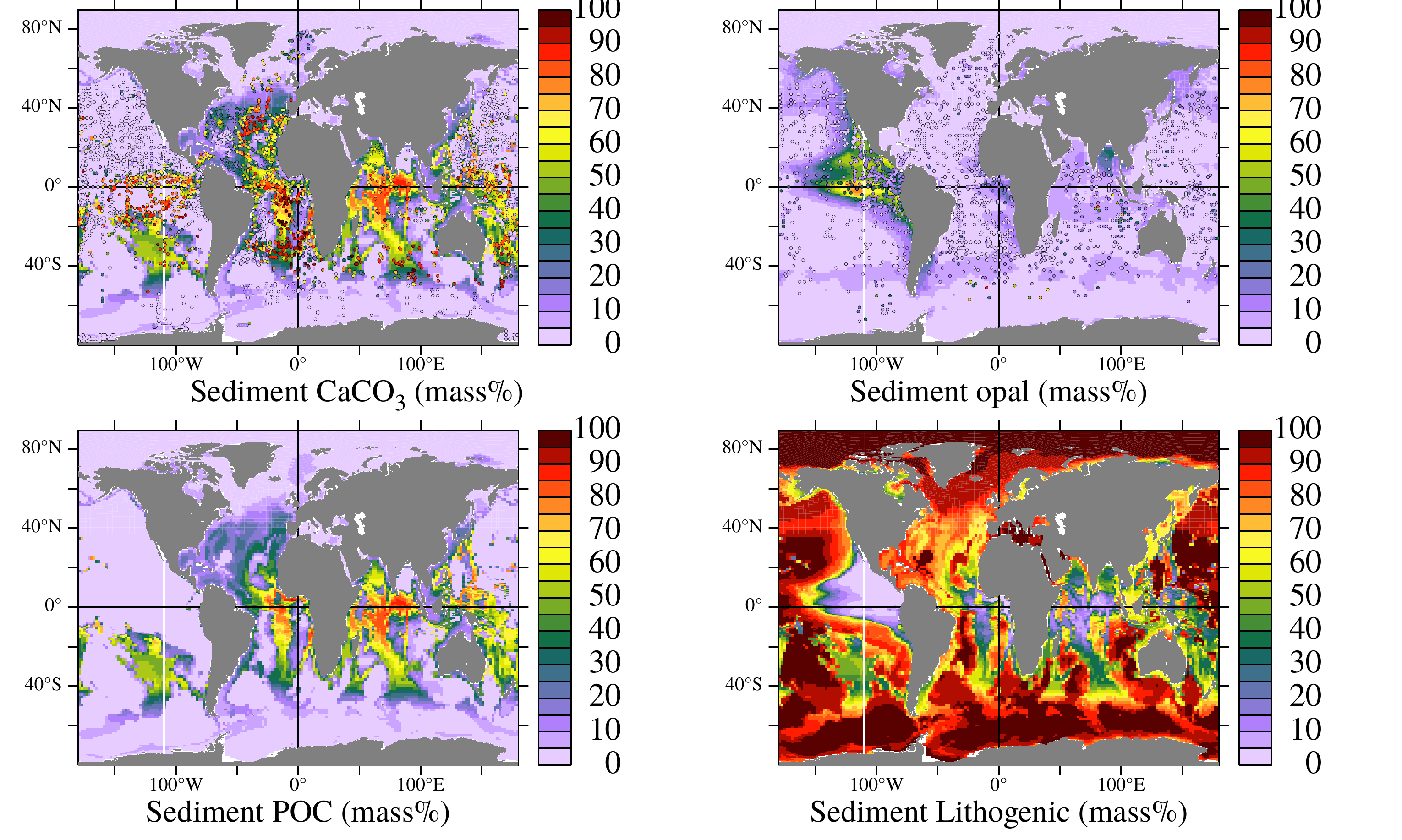}
    \caption{Particle composition of the sediment, average over depth and the
    last 1000\;yr of a 251\;kyr burst-coupled simulation.
    The calcium carbonate and biogenic silica data---plotted as discs on the
    same colour scale---are from \citet{archer1996} (revised dataset at
    \citet{data::archer1999}).}
    \label{fig:maps:particles}
\end{figure}

Figure~\ref{fig:maps:particles} presents the particle composition of the
sediment according to our model.


\begin{figure}
    \centering
    \includegraphics[width=.49\linewidth]{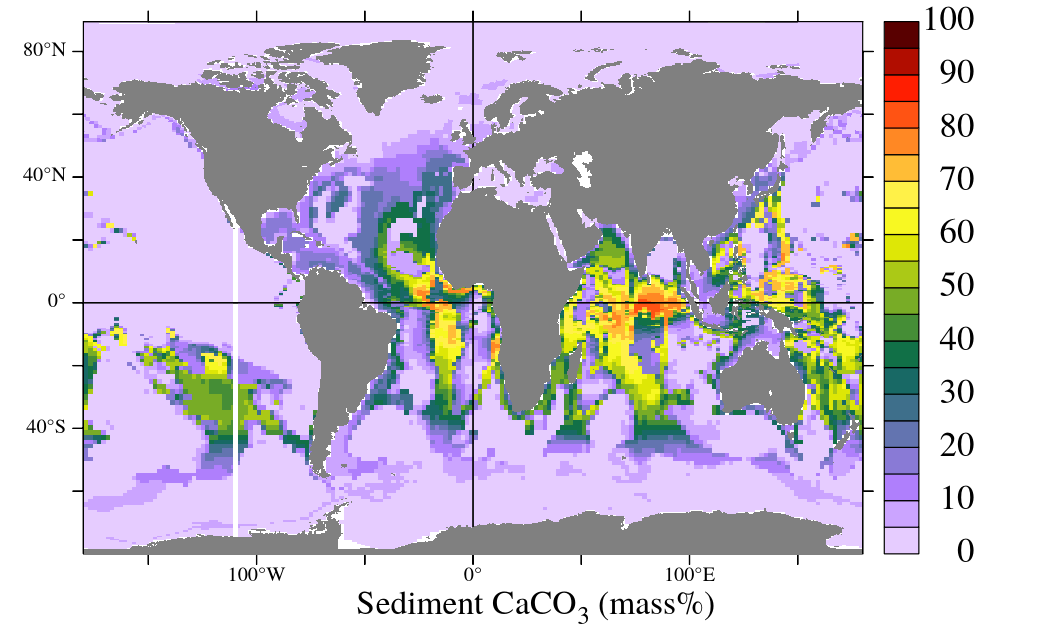}
    \includegraphics[width=.49\linewidth]{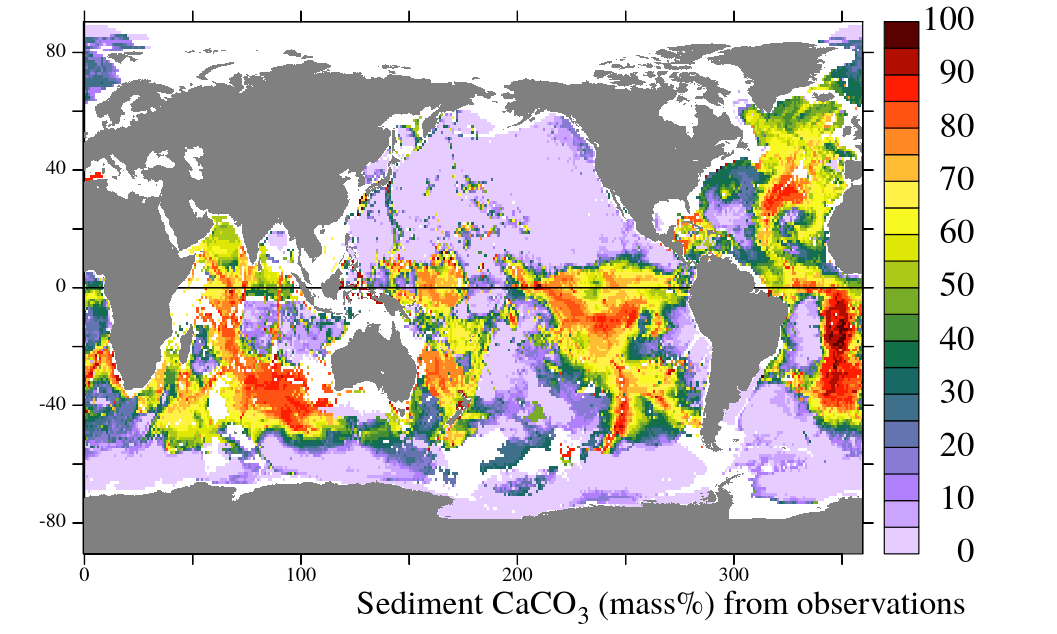}
    \caption{Mass fraction of calcium carbonate according to our model, and from
    coretop data \citep{archer1996}.}
    \label{fig:maps:CaCO3}
\end{figure}
In Figure~\ref{fig:maps:CaCO3}(b) observations by \citet{archer1996} are
presented.
The model underestimates \chem{CaCO_3} concentrations at most places, and
especially in the equatorial East Pacific.
The northern Indian Ocean is reasonably represented, not the southern part.
The low \chem{CaCO_3} fraction in the Southern Ocean, however, is well
represented.
This was expected as this is likely to rather contain opal instead of calcium
carbonate.
The simulation shows high concentrations of lithogenic material in especially
the Southern and Arctic Oceans.
Even though the opal belt is reproduced by the model
(Section~\ref{sec:pelagic} and \citet{tjiputra2020}), the
sedimentary biogenic silica fraction is nonetheless underestimated.

\begin{figure}
    \centering
    \includegraphics[width=\linewidth]{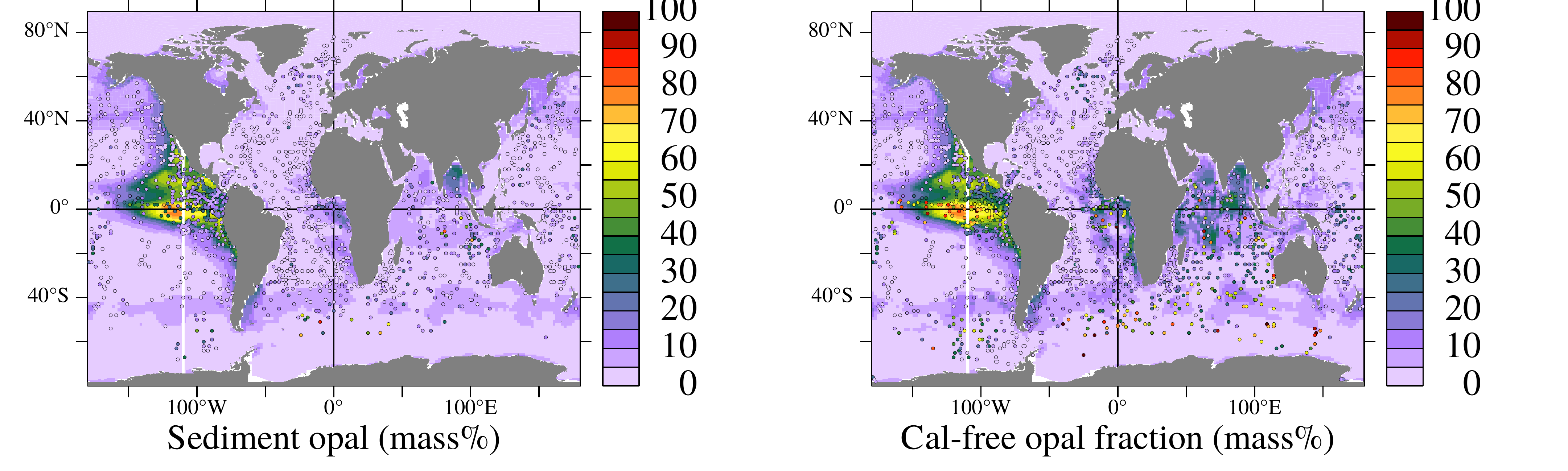}
    \caption{\chem{bSiO_2} fraction of the sediment, average over full model
    active sediment depth (0--15\;cm) and the
    last 1000\;yr of a 251\;kyr burst-coupled simulation.
    The biogenic silica data---plotted as discs on the same colour scale---are
    from \citet{archer1996} (revised dataset at \citet{data::archer1999}).
    \textbf{Left:} relative to total particulate; \textbf{right:} relative to
    particles excluding \chem{CaCO_3}.}
    \label{fig:maps:opal}
\end{figure}

\begin{figure}
    \centering
    \includegraphics[width=\linewidth]{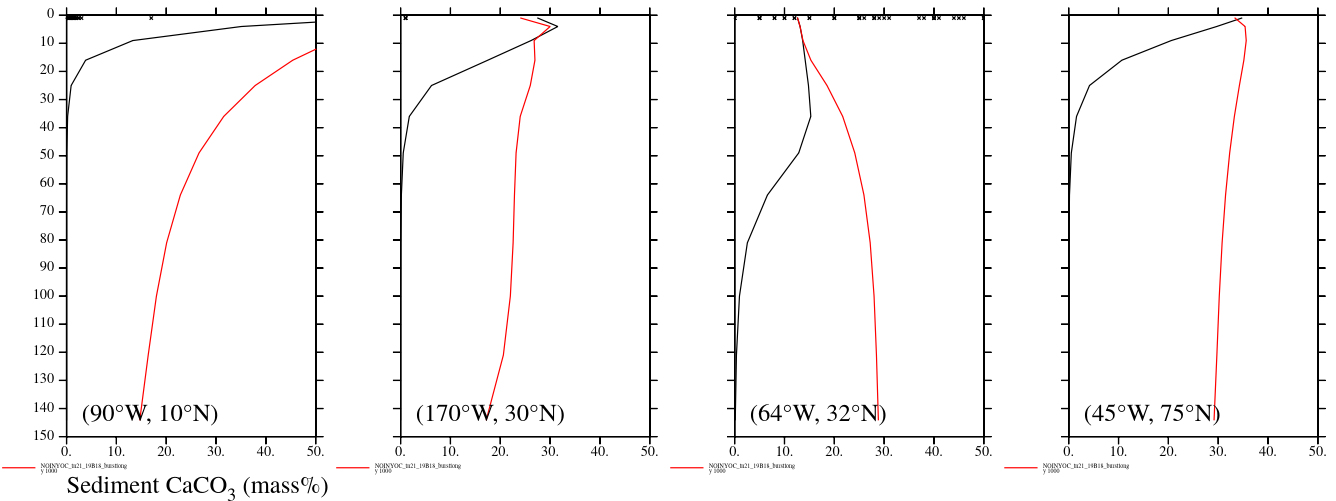}
    \caption{Profiles of \chem{CaCO_3} in the eastern equatorial Pacific Ocean,
    central North Pacific, BATS station and Weddell Sea.
    The vertical axis is the sediment depth (mm); the horizontal mass fraction
    of \chem{CaCO_3}.
    The black line is after 1000\;yr of normal BLOM/iHAMOCC simulation
    (\emph{Coupled});
    the red line is also 1000\;yr but every 250\;yr the sediment is decoupled
    to run 100\;kyr stand-alone, (\emph{BurstLong}).}
    \label{fig:profiles:calcite}
\end{figure}

\begin{figure}
    \centering
    \includegraphics[width=\linewidth]{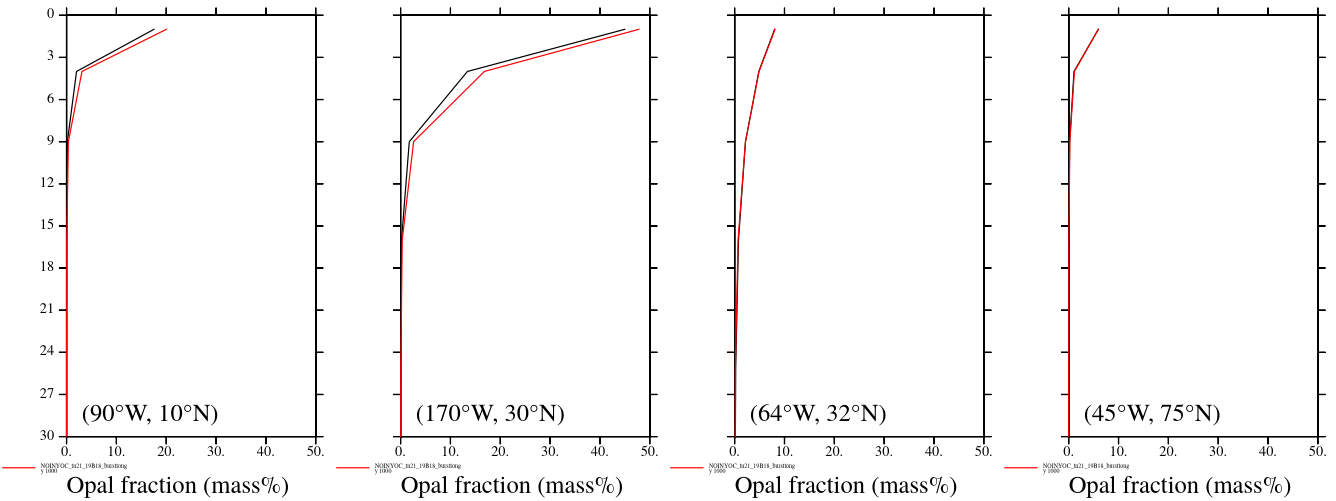}
    \caption{Profiles of b\chem{SiO_2} in the eastern equatorial Pacific Ocean,
    central North Pacific, BATS station and Weddell Sea.
    The vertical axis is the sediment depth (only upper 30\;mm); the horizontal
    mass fraction of b\chem{SiO_2}.
    The black line is after 1000\;yr of normal BLOM/iHAMOCC simulation
    (\emph{Coupled}); the red line is also 1000\;yr but every 250\;yr the
    sediment is decoupled to run 100\;kyr stand-alone, (\emph{BurstLong}).}
    \label{fig:profiles:bsi}
\end{figure}

\begin{figure}
    \centering
    \includegraphics[width=\linewidth]{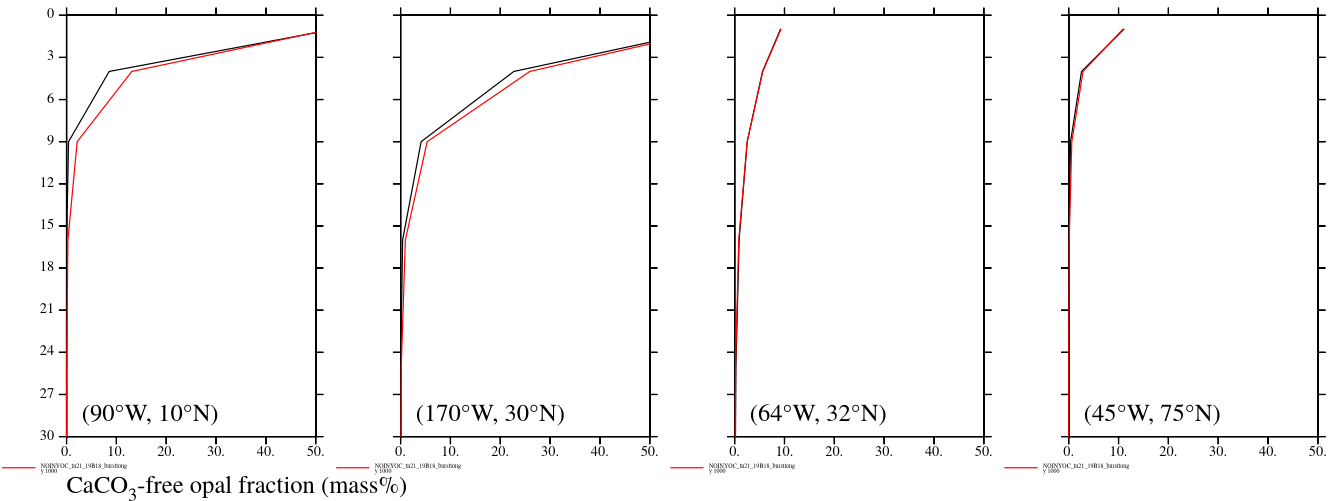}
    \caption{Profiles of the \textbf{calcium carbonate free} mass fraction of
    b\chem{SiO_2} in the eastern equatorial Pacific Ocean, central North
    Pacific, BATS station and Weddell Sea.
    The vertical axis is the sediment depth (only upper 30\;mm); the horizontal
    mass fraction of b\chem{SiO_2}.
    The black line is after 1000\;yr of normal BLOM/iHAMOCC simulation
    (\emph{Coupled}); the red line is also 1000\;yr but every 250\;yr the
    sediment is decoupled to run 100\;kyr stand-alone, (\emph{BurstLong}).}
    \label{fig:profiles:cfbsi}
\end{figure}


\begin{figure}
    \centering
    \includegraphics[width=.8\linewidth]{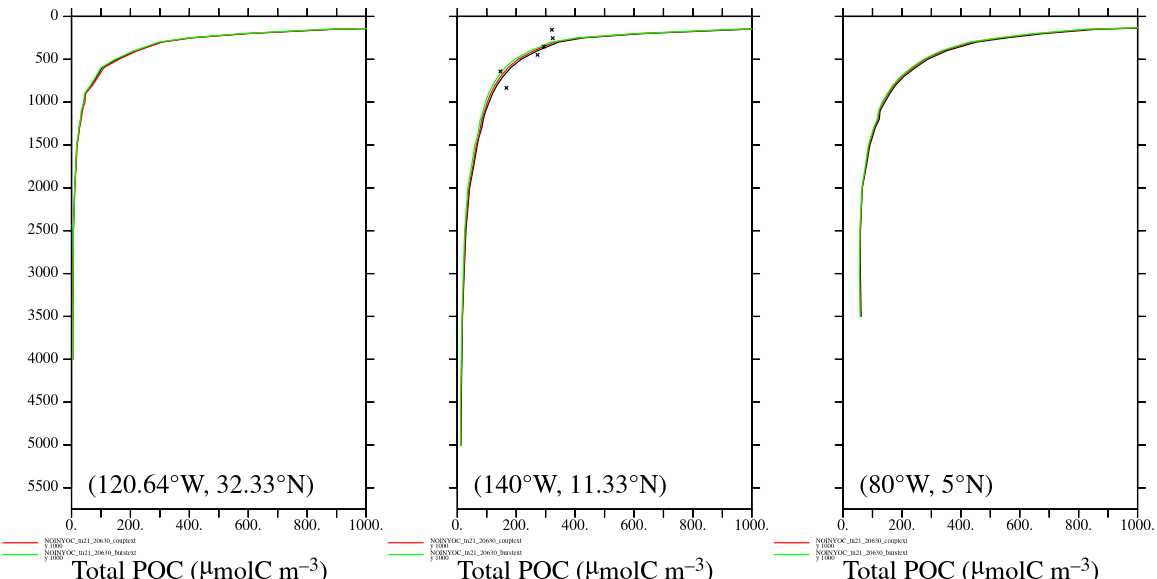}
    \caption{Profiles of the \chem{POC} concentration in the Pacific Ocean seawater.
    From left to right: the Patton Escarpment, MANOP Site S, and the Gulf of
    Panama.
    The vertical axis is the water column depth (m).
    The black line is after 1000\;yr of normal BLOM/iHAMOCC simulation
    (\emph{Coupled}); the red line after 1672 year; and the green line is after
    1000\;yr plus and additional 1000\;yr coupled burst coupled, interrupted by
    decoupled sediment simulations of run 50\;kyr stand-alone
    (\emph{BurstLong}).
    The crosses are measurements from \citet{lam2011}.}
    \label{fig:seawaterprofiles:poc}
\end{figure}

\begin{figure}
    \centering
    \includegraphics[width=.8\linewidth]{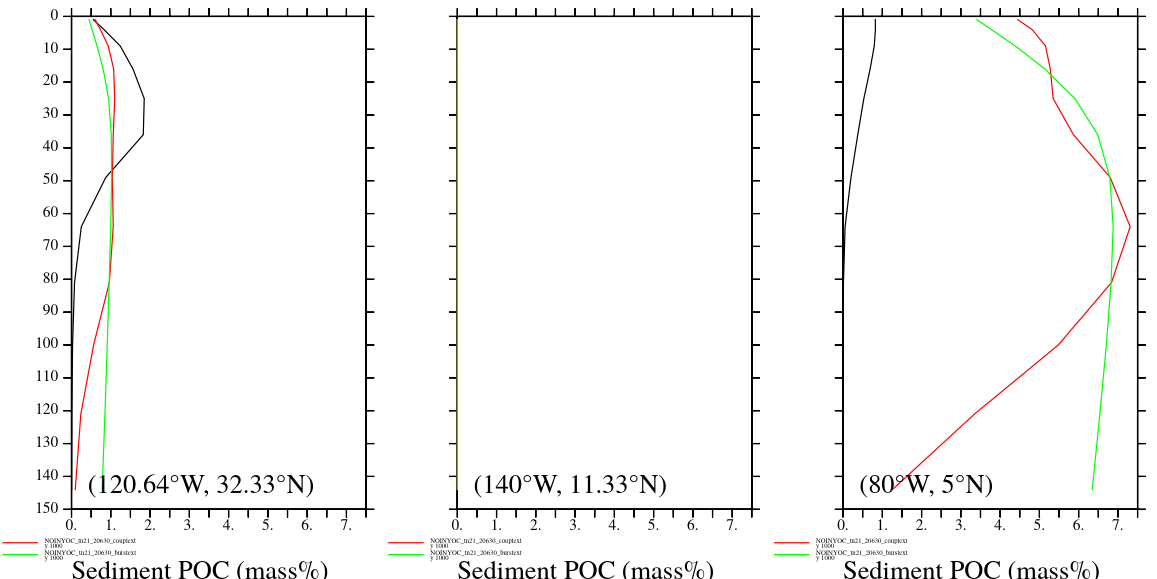}
    \caption{Profiles of the mass fraction of \chem{POC} in the Pacific Ocean.
    From left to right: the Patton Escarpment, MANOP Site S, and the Gulf of
    Panama.
    The vertical axis is the sediment depth (mm).
    The black line is after 1000\;yr of normal BLOM/iHAMOCC simulation
    (\emph{Coupled}); the red line after 1672 year; and the green line is after
    1000\;yr plus and additional 1000\;yr coupled burst coupled, interrupted by
    decoupled sediment simulations of run 50\;kyr stand-alone
    (\emph{BurstLong}).}
    \label{fig:profiles:poc}
\end{figure}

\begin{figure}
    \centering
    \includegraphics[width=.8\linewidth]{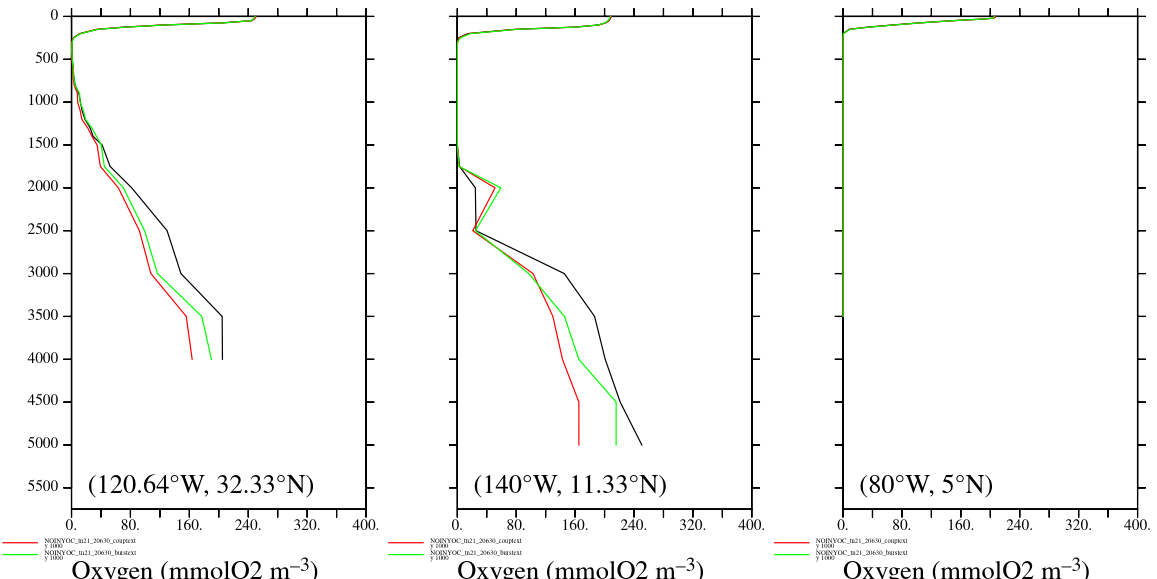}
    \caption{Profiles of the porewater concentration of \chem{O_2} in the
    Pacific Ocean seawater.
    From left to right: the Patton Escarpment, MANOP Site S, and the Gulf of
    Panama.
    The vertical axis is the water column depth in m.
    The black line is after 1000\;yr of normal BLOM/iHAMOCC simulation
    (\emph{Coupled}); the red line after 1672 year; and the green line is after
    1000\;yr plus and additional 1000\;yr coupled burst coupled, interrupted by
    decoupled sediment simulations of run 50\;kyr stand-alone
    (\emph{BurstLong}).}
    \label{fig:seawaterprofiles:o2}
\end{figure}

\begin{figure}
    \centering
    \includegraphics[width=.8\linewidth]{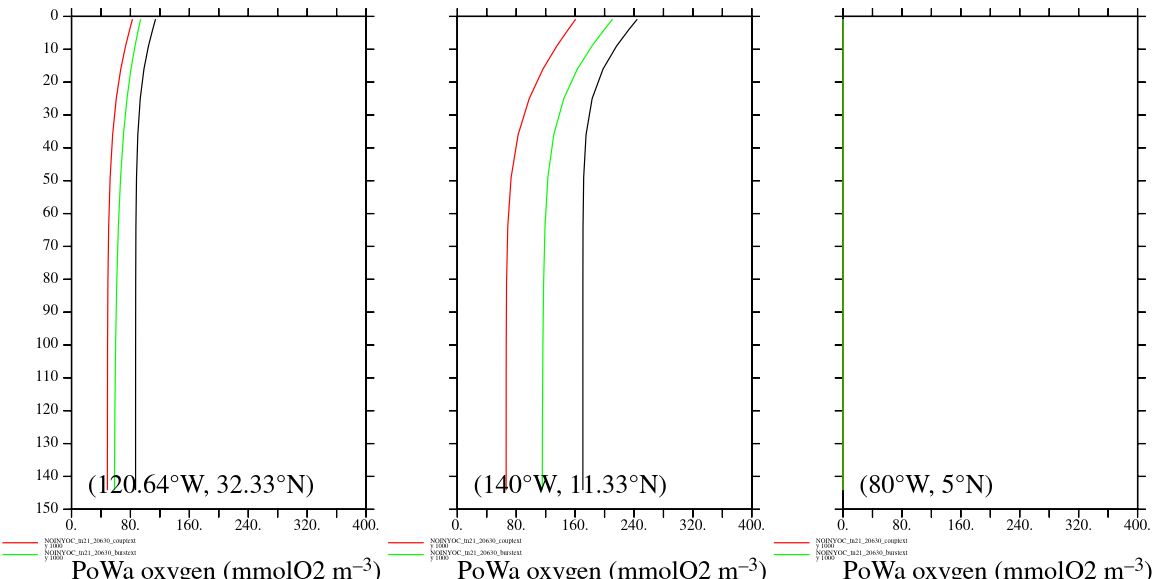}
    \caption{Profiles of the porewater concentration of \chem{O_2} in the
    Pacific Ocean.
    From left to right: the Patton Escarpment, MANOP Site S, and the Gulf of
    Panama.
    The vertical axis is the sediment depth in mm.
    The black line is after 1000\;yr of normal BLOM/iHAMOCC simulation
    (\emph{Coupled}); the red line after 1672 year; and the green line is after
    1000\;yr plus and additional 1000\;yr coupled burst coupled, interrupted by
    decoupled sediment simulations of run 50\;kyr stand-alone
    (\emph{BurstLong}).}
    \label{fig:profiles:o2}
\end{figure}

\begin{figure}
    \centering
    \includegraphics[width=\linewidth]{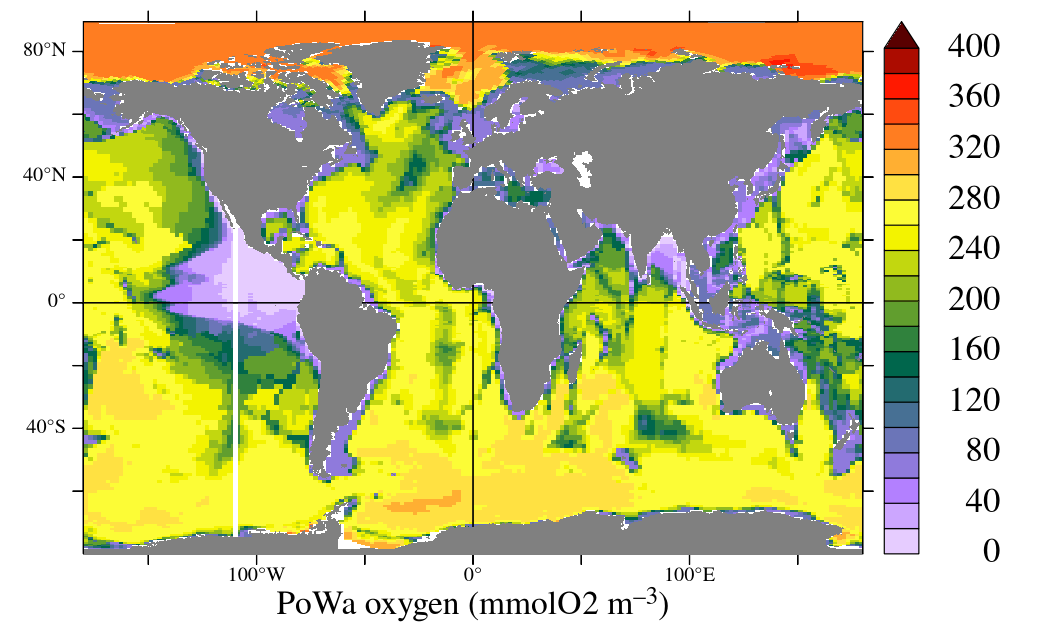}
    \caption{The depth-averaged porewater concentration of \chem{O_2} in the
    World Ocean.
    Yearly average at the end of \emph{BurstLong}.}
    \label{fig:map:o2}
\end{figure}

The \chem{POC} at the Patton Escarpment (Fig.~\ref{fig:profiles:poc}, left panel) has
the right asymptotic order of magnitude, 0.8\,\% compared with 1.4\,\% from
observations \citep[p.~231]{book::sarmiento2006}, but the shape is wrong
(starting at the surface below 0.5\,\% at the surface instead of 1.7\,\% in
observations).
The seawater profiles (Fig.~\ref{fig:seawaterprofiles:poc}) show some
consistency in that the highest concentration near the bottom are in the right
panel (Panama, where we expect high organic particle loads), consistent with the
about 6\,\% in the sediment.
Of course, percentages cannot be directly compared with concentrations; it
depends also on the other particles' sedimentation.
The zero value in the middle panel of the sediment profiles is inconsistent with
the particles in the seawater (latter compares well with observations).

The oxygen profiles in the seawater (Fig.~\ref{fig:seawaterprofiles:o2}) and the
sediment (Fig.~\ref{fig:profiles:o2}).
Oxygen concentrations are practically zero in the Gulf of Panama (see also the
map of Fig.~\ref{fig:map:o2}).
The sediment profiles have the right basic shape
\citep[p.~231]{book::sarmiento2006}, but they don't become zero at the Patton
Escarpment (continental margin, left panel of Fig.~\ref{fig:profiles:o2}).
This can of course be because of the too low resolution of the model to resolve
the continental margin.
The depth-average oxygen concentration in the sediment shows the hypoxic/anoxic 
areas around continents.


The burial rates from near the end of the \SI{251}{\kilo\year} simulation of
\emph{BurstLong} are given in Table~\ref{tab:burial}.

\begin{table}
\begin{tabular}{lSlll}
\toprule
Type of particle    & \multicolumn{4}{l}{Global burial rate}    \\
                    & Tmol/yr       & $\%_{>200m}$  & Tg/yr     & Tg/yr     \\
\midrule
Organic carbon      & 28 C          & 63            & 337 C     & 843 \chem{C_6H_{12}O_6} \\
Calcium carbonate   & 26 C          & 91            & 318 C     & 2649 \chem{CaCO3}       \\
Biogenic silica     & 23 Si         & 29            & 644 Si    & 1375 \chem{SiO2}       \\
Clay mineral        & 1.15          & 84            & 1.15      & 1.15                    \\
\bottomrule
\addlinespace
\end{tabular}
\caption{The burial rates after the \SI{251}{\kilo\year} simulation of
\emph{BurstLong}.}
\label{tab:burial}
\end{table}

Estimates of burial have a large variance.
For calcium carbonate the burial ranges typically from
\SIrange{100}{150}{\tera\gram\carbon\per\year} \citep{cartapanis2018}.
The model yields a burial that is more than two times higher than their upper
limit.
Then again, earlier estimates of \chem{CaCO_3} burial were higher and much
closer to our result \citep{milliman1993,milliman1996}.
In the model, when calcium carbonate is undersaturated, it dissolves following a
first-order dissolution.
The dissolution rate constant may be adjusted, and a higher-rate dissolution
parameterisation could be considered as well \citep[e.g.][]{subhas2015}.
Recent estimates of biogenic silica burial are
\sisetup{separate-uncertainty}%
\SI{202(115)}{\tera\gram\silicon\per\year} \citep{treguer2013}, whereas our
model yields a factor of three higher burial rate.
Also our carbon burial is overestimated with a factor of three.
Compare iHAMOCC's \SI{337}{\tera\gram\carbon\per\year} with
\SI{112}{\tera\gram\carbon\per\year} \citep{dunne2007} and
\SI{156}{\tera\gram\carbon\per\year} \citep{burdige2007}.

The coupled simulation (\emph{Coupled}) shows a relatively continuous rise of
particle content, because the sediment is filling up.
The burial is close to zero at the beginning of the simulation, because no
particles are present yet in the bottom sediment layer.
After burst coupling (\emph{BurstLong}), the sediment is spun up and fluxes are much
higher and without any obvious trend.

\subsection{Runtime efficiency}
Using the burst-coupling method we reached a model preformance increase per
model year of about a factor of~19.
This is on top of an about 20\,\% performance gain because of the refactoring
done to the code before and during implementation of the burst-coupling method.

\section{Discussion and conclusion}     \label{sec:discussion}
A complete steady state is only reached for \chem{CaCO_3} and PO\chem{C} far
beyond 100\,000\;yr, suggesting that (1) models that are similar in the relevant
respects need such a long spin-up and (2) the real ocean needs more than
100\,000\;yr to equilibrate.
In reality this may be shorter.

One issue with determining the equilibration time is the choice of the
coupling/decoupling procedure.
Instead of one parameter (time), it introduces three independent parameters,
namely coupled time, decoupled time and the number of iterations (or total
integration time).\footnote{One could deviate from our model scheme by allowing
variable coupled or decoupled periods, allowing for instance to change the
coupling state as soon as the relevant ocean component approaches a steady
state.
One might also use a climatology from one based on the last 50\;yr of a coupled
simulation; this should consider the extend of interyearly variablity.}
The 50\,000\;yr of stand-alone sediment integration is very long, chosen
for the practical reason that the sediment module integrates very fast, so why
not spin it up to a near steady state.
A consequence of our model setup is that the total integration time may not be a
good measure for how long an ocean model needs to get in whack.

Moreover, it appears that the full ocean system in \emph{BurstLong} is not yet
quite in whack and needs at least a couple more iterations.
This is, abducing from the previous paragraph, due to too little coupled
integration time.

Our model underestimates the sediment calcium carbonate mass fraction,
especially in the equatorial Pacific (Fig~\ref{fig:maps:particles}
or~\ref{fig:maps:CaCO3}).
The general distribution patterns of calcium carbonate are realistic; they
relate to the topography as the observations.
The biogenic silica is underestimated in the opal belt and overestimated in the
equatorial Pacific.
It seems that in the model diatoms rule in the equatorial Pacific, but in
reality coccolithophores are important there.
Also carbon burial is likely strongly overestimated and hence pressing the
calcium carbonate mass fraction burial.

Overall burial of all particles in iHAMOCC could be increased by increasing the
bioturbation coefficient that is not well known.
This would speed up the burial process
\citep[e.g.][pp.~41--44]{book::boudreau1997}.
It is not know if the bioturbation of \SI{1e-9}{\metre\squared\per\second} that
we used is realistic.

\section{Code availability}
The sediment burst-coupling code may be obtained at
\url{https://puszcza.gnu.org.ua/projects/burst-coupling/}.
For this study, different revisions of the code from
\texttt{105:fe270f4aa7db} to \texttt{121:dad56c57eb86} were used.
The repository also contains a
\href{https://hg.gnu.org.ua/hgweb/burst-coupling/file/c2feaf25645d/howto_sediment.pdf}{manual}
for using burst coupling with BLOM/iHAMOCC\@.
The mathematical variable names correspond to the Fortran variables as described
in Table~\ref{tab:variables} in the appendix.

Analysis scripts can be found at \url{https://hg.sr.ht/~marco/sediment-analysis}.
A short manual for using these scripts can be found in Appendix~\ref{app:post}.
Model output is available on request.

\section{Authors' contributions}
The model and the simulations were designed by MvH, JS, JT and CH\@.
MvH designed the burst coupling code structure with contributions from JS, JT,
CH, MB and AG\@.
The manuscript was prepared by MvH with close collaboration and major
contributions from CH, JS, and JT\@.

The authors declare that they have no conflict of interest.

\begin{acknowledgements}
We would like to thank Mats Bentsen for the useful discussions about the model
design.
We thank Alok Gupta and Anne Mor\'ee for helping with the model setup.

This study was supported by the project ``Overturning circulation and its
implications for the global carbon cycle in coupled models'' (ORGANIC, The
Research Council of Norway, grant No.~239965).
This work was also supported through project CRESCENDO (Coordinated
Research in Earth Systems and Climate: Experiments,
Knowledge, Dissemination and Outreach; Horizon 2020 European
Union's Framework Programme for Research and Innovation, grant
No.~641816, European Commission).

The authors wish to acknowledge the use of
\href{https://ferret.pmel.noaa.gov/Ferret/}{Ferret}, a product of
\href{http://www.noaa.gov/}{NOAA}'s Pacific Marine Environmental Laboratory.
The plots in this paper were created by the Ferret visualisation library
\href{http://www.nongnu.org/complot/}{ComPlot} \citep{vanhulten2017:complot} and
\href{http://www.gnuplot.info/}{gnuplot}.
\end{acknowledgements}

%
\DeclareRobustCommand{\DutchName}[4]{#1,~#3~#4}

\bibliography{clim_books,clim_data,clim_papers,clim_preprints,clim_software,clim_theses}{}
\bibliographystyle{copernicus}

\onecolumn
\clearpage
\appendix

\section{Model variables}               \label{app:model}

\begin{table}[h]
\centering
\begin{tabular}{>{\(}l<{\)}lrl>{\tt}l}
\toprule
\mathrm{Symbol}     & Description                       & Value         & Unit                              & \rm Code  \\
\midrule
\multicolumn{5}{l}{\textit{state variables of the solid fraction ($s$)}} \\
c_s                 & Concentration of solid sediment component $s$ & Variable & concentration                      & sedlay(:,:,:,:) \\
\addlinespace
\chem{OC}           & Organic carbon                    & Variable      & \si{\mol\phosphorus\per\cubic\deci\metre} & sedlay(:,:,:,issso12) \\
\chem{CaCO_3}       & Calcium carbonate                 & Variable      & \si{\mol\carbon\per\cubic\deci\metre}     & sedlay(:,:,:,isssc12) \\
\chem{bSiO_2}       & Biogenic silica                   & Variable      & \si{\mol\silicon\per\cubic\deci\metre}    & sedlay(:,:,:,issssil) \\
\chem{clay}         & lithogenic (via dust)             & Variable      & \si{\kilo\gram\per\cubic\metre}           & sedlay(:,:,:,issster) \\
\midrule
\multicolumn{5}{l}{\textit{state variables of the porewater ($d$)}} \\
c_d                 & Concentration of dissolved porewater component $d$ & Variable & molar concentration           & powtra(:,:,:,:) \\
\addlinespace
\chem{DIC}          & Dissolved inorganic carbon        & Variable      & \si{\mol\carbon\per\cubic\deci\metre}     & powtra(:,:,:,ipowaic) \\
A_T                 & Total alkalinity                  & Variable      & \si{\equiv\per\cubic\deci\metre}          & powtra(:,:,:,ipowaal) \\
\chem{PO_4}         & Phosphate                         & Variable      & \si{\mol\phosphorus\per\cubic\deci\metre} & powtra(:,:,:,ipowaph) \\
\chem{O_2}          & Oxygen                            & Variable      & \si{\mol\oxygen\per\cubic\deci\metre}     & powtra(:,:,:,ipowaox) \\
\chem{N_2}          & Molecular nitrogen                & Variable      & \si{\mole\nitrogen\per\cubic\deci\metre}  & powtra(:,:,:,ipown2) \\
\chem{NO_3}         & Nitrate                           & Variable      & \si{\mol\nitrogen\per\cubic\deci\metre}   & powtra(:,:,:,ipowno3) \\
\chem{Si(OH)_4}     & Silicic acid                      & Variable      & \si{\mol\silicon\per\cubic\deci\metre}    & powtra(:,:,:,ipowasi) \\
\midrule
\multicolumn{5}{l}{\textit{parameters}} \\
v_{s\rightarrow d}  & Dissolution reaction rate         & Variable      & \si{\kilo\mole\per\cubic\metre\per\second} & \\
\rho                & Bulk sediment density             & Variable      & \si{\kilo\gram\per\cubic\metre}   & sedlo \\
\rho_\chem{CaCO_3}  & Density of \chem{CaCO_3}          & 2600          & \si{\kilo\gram\per\cubic\metre}   & calcdens \\
\rho_\chem{bSiO_2}  & Density of \chem{bSiO_2}          & 2200          & \si{\kilo\gram\per\cubic\metre}   & opaldens \\
\rho_\chem{OC}      & Density of organic carbon         & 1000          & \si{\kilo\gram\per\cubic\metre}   & orgdens \\
\rho_\chem{clay}    & Density of clay (quartz)          & 2600          & \si{\kilo\gram\per\cubic\metre}   & claydens \\
M_\chem{CaCO_3}     & Molecular weight of \chem{CaCO_3} &  100          & \si{\kilo\gram\per\kilo\mole}     & calcwei \\
M_\chem{bSiO_2}     & Molecular weight of \chem{bSiO_2} &   60          & \si{\kilo\gram\per\kilo\mole}     & opalwei \\
M_\chem{OC}         & Molecular weight of organic carbon & 100          & \si{\kilo\gram\per\kilo\mole}     & orgwei \\
R                   & Stoichiometric ratio \chem{C:P}   & 122           & --                                & rcar \\
R                   & Stoichiometric ratio \chem{N:P}   & 16            & --                                & rnit, rno3 \\
R                   & Stoichiometric ratio \chem{-O_2:P} & 172          & --                                & ro2ut \\
\mathcal{B}         & Bioturbation diffusion coefficient & \num{1.e-9}  & \si{\metre\squared\per\second}    & sedict \\
\mathcal{D}         & Porewater diffusion coefficient   & \num{1.e-9}   & \si{\metre\squared\per\second}    & sedict \\
\phi                & Porosity                          & \numrange{0.62}{0.85} & --                        & porwat(:) \\
V_\mathrm{f}        & Total solid sediment volume       & $\lessapprox 1-\phi$ & --                         & solfu \\
\bottomrule
\addlinespace
\end{tabular}
\caption{The sediment model's parameters and variables with values, associated
units, and variable names in the Fortran code of BLOM/iHAMOCC.}
\label{tab:variables}
\end{table}

\section{Postprocessing and analysis}   \label{app:post}

Here is shown how the model output is processed and analysed.
Besides for reproduction of the final results, the reader may freely use any or
all of these methods and code for their own purpose, though most of it is quite
specific for BLOM/iHAMOCC output.

The pipelines are partial, or partially automated if you will.
The scripts depend on \href{https://code.mpimet.mpg.de/projects/cdo}{CDO} and
\href{https://ferret.pmel.noaa.gov/Ferret/}{Ferret}.
The relevant scripts can be accessed at
\url{https://hg.sr.ht/~marco/sediment-analysis}.
After running the model, a model output file must be given as input to
\texttt{process4analysis}, and an output directory can optionally be specified
as well:
\begin{verbatim}
process4analysis -i model-output.nc -o processed
\end{verbatim}
This pipeline ends with an \texttt{out.nc} in your output directory.
Go into that directory and run \texttt{plot-tracers.gnuplot}, e.g.\ with the
\texttt{load} command on the Gnuplot prompt, to make timeseries plots.

On your system you may need to remove or modify the \texttt{module load}
commands.
Note that strftime(3) as implemented in the GNU C Library can handle years
beyond 9999, which may be useful for long simulations.
For six digits you can use a conversion specification of \texttt{\%06Y}.
This is not part of BSD~libc or the ISO C standard.

\enlargethispage{2cm}
For maps in either the seawater or the sediment or transsects, one may consider
using \href{http://www.nongnu.org/complot/}{ComPlot}; full documentation is
included in that package.

\section{Recommendations}               \label{app:recommendations}

The development, implementation, simulation and analysis of this burst-coupling
method took a much longer time than expected.
This was due to a combination of several reasons, but here the main author will give
only the technical (and perhaps philosophical) elements that played a role.
This could be useful for those who consider doing something similar.

The climate system is complex and it is only a wonder that we mortals are able to
capture some of its complexity in a quantitative simulation model.
As a consequence, climate models are bound to be complex as well; otherwise they
would not be a good representation of reality.
Of course, they do not need to be close to a complete representation of reality,
because models of low complexity that describe only one or a small subset of
mechanisms can be very useful to investigate those mechanisms.
However, here I talk about models of high complexity that are created to be used
for prognostic purposes (e.g.\ CMIP projections).

The model that is used in this study is BLOM/iHAMOCC, which is part of NorESM.
BLOM/iHAMOCC is developed to be used as part of CMIP projections and it is
always the question, and it should be considered, if one should use it for
other studies as well.\footnote{There is also a discussion if and in what way
the many model simulations as part of CMIP are useful, but the answer to that is
far from trivial and I will not discuss that here.}
In the case of this study this is probably waranted, because of the opportunity
to run the model efficiently to a steady state, which is a useful starting point
for almost any simulation including for the CMIP\@.
The sediment component of the model is in fact suitable for the long-term
simulations, because it is fast whilst still of sufficient complexity for the
purposes here.

That said, the motivation to use this model is politically driven.
Every nation state that feels important must run their own model, or so the
emergent driving power of the descientised enterprise thinks.
This is a waste of resources, both computational and people-wise.
It is better to choose a model or method based on more scientific grounds.

Again, the way BLOM/iHAMOCC has been used in this study is not particularly bad.
However, its coding quality is not very high.
It is at least behind at least some other models.
One example is \href{https://www.nemo-ocean.eu/}{NEMO}\@.\footnote{The main
author is familiar only with the intrinsics of NEMO and BLOM/iHAMOCC, but all
statements made in these recommendations can be made based on only those two.}
This statement is of course not consistent with \emph{model democracy},
something that is often presumed but never really tested.
Of course there are model intercomparison studies and the models get evaluated
(visually, goodness-of-fit and so), which shows that some models are much better
than others, but the idea of model democracy persists.

Besides the usual way of evaluating through one or more goodness-of-fits,
another notion of reliability is generally ignored throughout the literature
\citep[e.g.][\S{}2.4]{thesis::vanhulten2014}.
In this second sense, NorESM and at least some of its components, notably
BLOM/iHAMOCC, are not reliable.
This becomes clear from even superfacial code and code structure inspection.
Both the efforts that came from this study as well as more recent developments
\citep{tjiputra2020} have improved this a lot, but there is still a lot to be
desired.
In this sense it will always be behind certain other models like NEMO.
There are good reasons for this.
The modelling team of the ``French'' model NEMO (and the larger IPSL model and
even the PISCES biogeochemical subcomponent of NEMO) is larger.
This should be taken at face value and as such the choice of using BLOM/iHAMOCC
was possibly the wrong one.

That said, NEMO does not have a proper sediment component (there is something
but it is not recently tested or well developed).
That is a motivation to use another model or implement a sediment model.
Nonetheless, it is not sufficient reason to use BLOM/iHAMOCC for this.

One of the difficulties with implementing the burst-coupling method into
BLOM/iHAMOCC was that the code was not well written.
It had little consistency and its structure is hard to follow.
After the main author had refactored and debugged a part of the code, the model
ran about 20\,\% faster in coupled mode.
Some
\href{https://github.com/NorESMhub/BLOM/blob/master/doc/coding-guidelines.rst}{guidelines were adopted}
and used thoughout the model code.
This can be seen by browsing through the commits of the last year or the last
couple of years.
The way of developing the model as a community has improved as well.
This is largely due to its availability through a public repository and that
it is published under a free software license.
At the same time, culture change is slow and a stronger DevOps philosophy could
strengthen the model's development further.
The recent coding efforts made the code undoubtly more readable, less buggy and
more efficient.
In the meantime, other models have improved as well.
Whether to continue using NorESM for CMIP simulations, and whether to use this
model and BLOM/iHAMOCC for a specific study, should be carefully considered and
also evaluated during development.

If it is decided to use NorESM and BLOM/iHAMOCC for further studies, I would
recommend introducing the latest and best version of the burst-coupling code in
a more current, stable version of NorESM.
The long burst simulation should then be repeated and an evaluation of the
sediment and reevaluation of the seawater should be performed.
\end{document}